\documentclass[aps,prb,notitlepage,twocolumn,noeprint]{revtex4-2} 

\usepackage{amsmath,amssymb}
\usepackage{graphicx,epstopdf}
\usepackage{dcolumn}
\usepackage{verbatim, float}
\usepackage{xcolor}
\usepackage[colorlinks=true]{hyperref}
\usepackage[normalem]{ulem}

\newcommand{\bdsym}{\boldsymbol}

\begin{document}
\title{Intrinsic Nernst effect of the magnon orbital moment in a honeycomb ferromagnet}
\author{Daehyeon An}
\email{daehyeon.an@kaist.ac.kr}
\affiliation{Department of Physics, Korea Advanced Institute of Science and Technology, Daejeon 34141, Republic of Korea}
\author{Se Kwon Kim}
\email{sekwonkim@kaist.ac.kr}
\affiliation{Department of Physics, Korea Advanced Institute of Science and Technology, Daejeon 34141, Republic of Korea}
\date{\today}

\begin{abstract}
The Nernst effect of the magnon orbital moment is theoretically investigated in a honeycomb ferromagnet, whose Hamiltonian contains the Heisenberg exchange, the Dzyaloshinskii-Moriya, the Kitaev, and the Zeeman interactions. More specifically, we obtain the magnon band structure, the Berry curvature, the magnon orbital moment Berry curvature, and the magnon orbital moment Nernst conductivity (MOMNC) for the Hamiltonian within the linear spin-wave theory in the polarized phase.
We found the magnon orbital moment Berry curvature is largely independent of the sign of the Chern number of the band.
For an experimental proposal, we estimate MOMNC of $\mathrm{CrI}_{3}$ for the Heisenberg-Kitaev-$\Gamma$-Zeeman model and Heisenberg-Dzyaloshinskii-Moriya-Zeeman model for a perpendicular external magnetic field. We found that certain Kitaev materials provide us with a broad tunability of the MOMNC through a magnetic field.
We envision that our findings lead to further investigations of the novel transport properties of the magnon orbital moment.
\end{abstract}
\maketitle

\section{Introduction}
Spintronics, or spin-transport electronics, is a research field investigating various ways to process, store, and detect information by exploiting the spin degree of freedom of the electron in solid state systems~\cite{Wolf2001Science.294.1488, Zutic2004RevModPhys.76.323}.
It is actively researched for and driven by technical applications, e.g. magnetic random access memory~\cite{Wolf2001Science.294.1488, Puebla2020CommunMater.1.24}, and racetrack memory~\cite{Parkin2008Science.320.190, Puebla2020CommunMater.1.24}, also due to fundamental spin physics therein, e.g. spin Hall effect~\cite{Hirsch1999PhysRevLett.83.1834,Kato2004Science.306.5703}, spin Seebeck effect~\cite{Seki2015PhysRevLett115.266601,Rezende2016PhysRevB93.014425,Uchida2010ApplPhysLett.97.172505}, and spin Hanle effect~\cite{Appelbaum2007Nature.447.7142, Lou2007NatPhys.3.197}. The orbital momentum of a charged particle contributes to the magnetic moment as the spin does.
The orbital momentum of electrons in the solid-state system has been defined in the modern theory~\cite{Xiao2010RevModPhys.82.1959} by Wannier function description~\cite{Thonhauser2005PhysRevLett.95.137205} and semiclassical dynamics~\cite{Xiao2005PhysRevLett.95.137204}.
Researchers have recently started to study exploiting the orbital degree of freedom of the electron in addition to spin and charge~\cite{Go2021EurophysLett.135.37001}, spawning a new field called orbitronics. For example, the orbital Hall effect has been identified as the root of the spin Hall effect in the centrosymmetric normal metal~\cite{Jo2018PhysRevB.98.214405} and anomalous Hall effects through spin-orbit coupling~\cite{Kontani2009PhysRevLett.102.016601}.
The orbital Hall effect is a universal phenomenon that appears in generic solids even if the orbital moment is quenched~\cite{Go2018PhysRevLett.121.086602}.

\begin{figure}[!htp]
    \includegraphics[width=240pt]{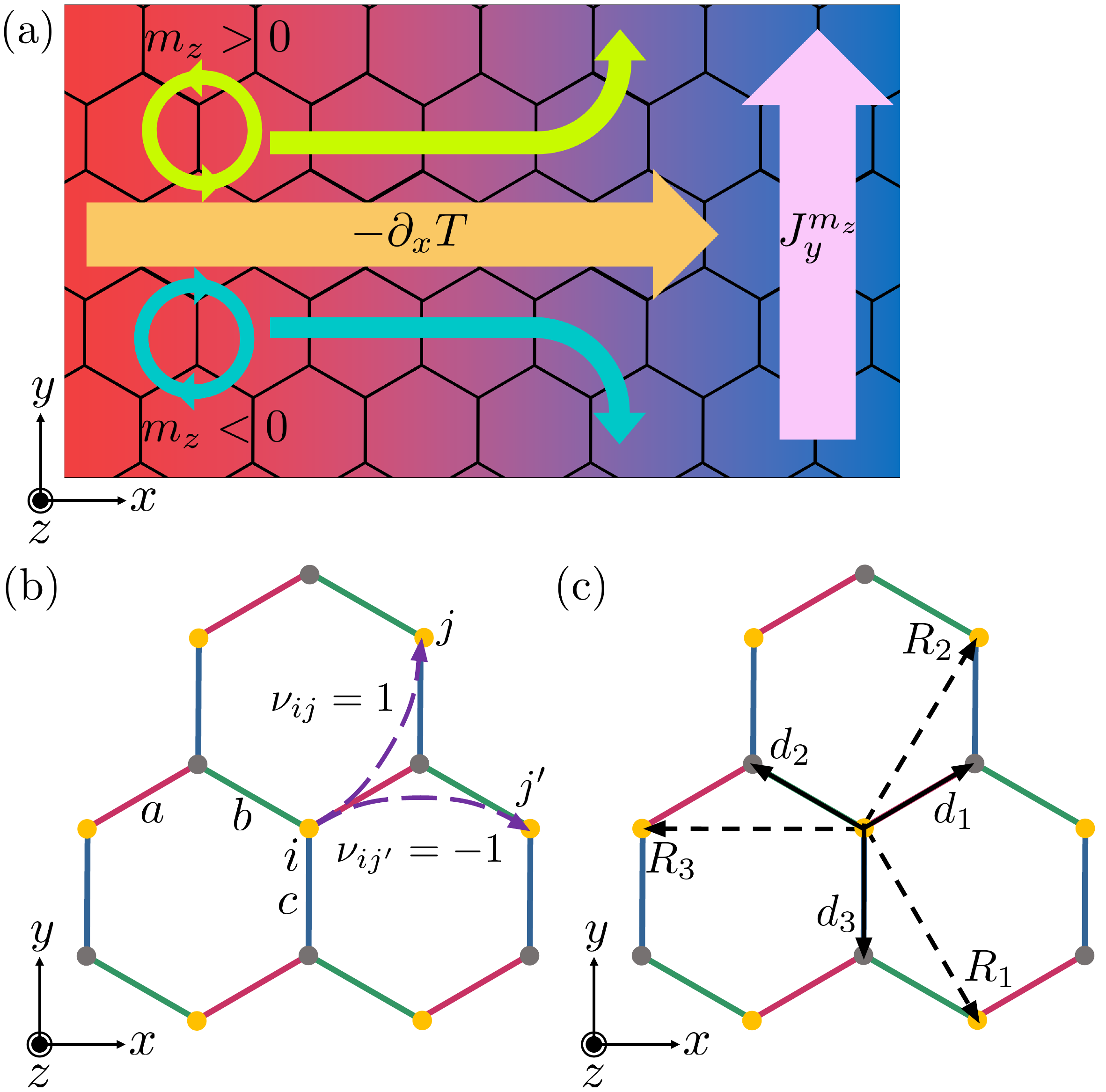} 
    \caption{
    (a) The schematic picture of the magnon orbital moment Nernst effect. The temperature gradient along the $x$-axis $-\partial_x T$ induces the magnon orbital moment current density $J^{m_z}_{y}$ flowing along the $y$ direction carrying the $z$ component magnon orbital moment, $m_z$. 
    (b) Yellow and gray dots are the $A$ and $B$ sublattices of the honeycomb structure, the signs $\nu_{ij}$ of DM interaction differ by the chirality shown as the dashed purple arrows, and the colors of bonds denote three types of the Kitaev interaction along the bonds, where red, green, blue are $a$, $b$, $c$ bond types of the interaction respectively.
    (c) The $\bdsym{d}$'s (solid lines) and $\bdsym{R}$'s (dotted lines) are the position vectors from the sublattice $A$ to the nearest neighborhoods and to the next-nearest neighborhoods respectively.
    }
    \label{fig1}
\end{figure}

A magnon is a quasiparticle of spin waves, which has spin but does not have any charge.
Magnonics, the magnon counterpart of spintronics~\cite{Kruglyak2010JPhysD43.264001, Chumak2015NatPhys.11.453, Bozhko2020LowTempPhys.46.383, Serga2010JPhysD.43.264002}, offers Joule heating-free spin-based information technology~\cite{Chumak2014NatCommun.5.4700, Chumak2015NatPhys.11.453} enabled by, e.g., long-distance propagation with low damping~\cite{Serga2010JPhysD.43.264002,Cornelissen2015NatPhys.11.1022}, and the terahertz signal generation and detection~\cite{Kampfrath2011NatPhoton.5.31, Li2020Nature.578.70}.
Considering the advancement of magnonics, it is natural to think about the magnon counterpart of orbitronics, which we dub magnon orbitronics.
Since the magnon has no charge and no natural mass, there are different physical definitions of the orbital-related quantity.
For example, the magnon orbital magnetic moment $\bdsym{\mu}^{O}_{n,\bdsym{k}}$ has been defined in Ref.~\cite{Neumann2020PhysRevLett.125.117209} by the orbital contribution of the magnetic moment of the magnon.
It is the response of the magnon energy to the magnetic field subtracting the spin contribution from the Zeeman effect, which is obtained by taking the derivative of the magnon energy with respect to the magnetic field and subtracting the spin of the magnon that comes from the Zeeman effect.
Also, the magnon angular momentum is defined in Refs.~\cite{Fishman2022PhysRevLett.129.167202,Fishman2022JCdsMaterPhys.51.015801,Fishman2023PhysRevB.107.214434} as the angular momentum $\hat{\bdsym{L}}=\hat{\bdsym{r}}\times\hat{\bdsym{p}}$ possessed by the magnon, where itself is gauge dependent so that the azimuthal angle integrated-quantity is suggested as a gauge-invariant quantity related to the angular momentum~\cite{Fishman2023PhysRevB.107.214434}.
Recently, in the paper by Go et al.~\cite{Go2024NanoLett.24.5968}, the magnon orbital moment is defined by $\hat{\bdsym{m}}\sim\hat{\bdsym{r}}\times\hat{\bdsym{v}}$, which captures the self-rotation of the magnon wave-packet, by adopting the modern theory of orbital magnetization which introduced the gauge independent orbital moment of the electron~\cite{Xiao2010RevModPhys.82.1959}.
Based on this, the magnon orbital moment Nernst effect is predicted in a honeycomb antiferromagnet without Dzyaloshinskii-Moriya (DM) interaction~\cite{Go2024NanoLett.24.5968}.
The work has been extended to the case with DM interaction in Ref.~\cite{To2024ArXiv.16004}.
Our work focuses on the magnon orbital moment $\hat{\bdsym{m}}\sim\hat{\bdsym{r}}\times\hat{\bdsym{v}}$ defined in Ref.~\cite{Go2024NanoLett.24.5968}.

The exact solution of the quasi-particle has been obtained in a honeycomb ferromagnet with the so-called Kitaev interactions~\cite{Kitaev2006AnnPhys.321.2}, which emerges in $4d$, $5d$ transition metal compound with strong spin-orbit coupling~\cite{Jackeli2009PhysRevLett.102.017205}.
In addition, in honeycomb spin systems, the DM interaction is allowed by symmetry considerations~\cite{Kim2016PhysRevLett.117.227201,Chen2018PhysRevX.8.041028}.
Both interactions, the Kitaev interactions and the DM interactions, affect the magnon band structures of a honeycomb ferromagnet such as $\mathrm{CrI}_{3}$, but it has been challenging to disentangle the effects of the two in experimental results~\cite{Chen2020PhysRevB.101.134418, Lee2020PhysRevLett.124.017201, Chen2021PhysRevX.11.031047, Zhang2021PhysRevB.103.134414, Brehm2024ArXiv.15964}.
In this paper, we investigate the Nernst effect of the magnon orbital moment $\hat{\bdsym{m}}$ in a honeycomb ferromagnet within the linear spin-wave theory and compare the effects of the Kitaev interaction and DM interaction. 

The rest of the paper is organized as follows. 
In Sec.~\ref{sect 2}, we explain our ferromagnetic system with the Heisenberg, DM, Kitaev, and Zeeman interaction. We obtain magnon dispersion based on the linear spin-wave theory where the system is in the fully spin-polarized phase. We analyze the energy bands and Berry curvature.
Using the Kubo formula, we obtain the magnon orbital moment Nernst conductivity (MOMNC).
In Sec.~\ref{sect 3}, we evaluate the MOMNC of $\mathrm{CrI}_{3}$ for the Heisenberg-Dzyaloshinskii-Moriya-Zeeman model, from now on $J$-$DM$-$B$ model, and the Heisenberg-Kitaev-$\Gamma$-Zeeman model, from now on $J$-$K$-$\Gamma$-$B$ model~\cite{Chen2021PhysRevX.11.031047}.
We show that both models exhibit the magnon orbital moment Nernst effect when $D$ and $K$ are finite, respectively and compare the two cases.
In Sec.~\ref{sect 4}, we conclude the paper by summarizing the results.

\section{Formalism}\label{sect 2}
In this section, we present our formalism to derive the MOMNC by applying the linear spin-wave approximation from the fully spin-polarized phase in the honeycomb ferromagnets, whose Hamiltonian contains the Heisenberg exchange, the DM, Kitaev, and Zeeman interaction.

\subsection{Model}

The system that we consider is modeled by
\begin{align}\label{H}
    H =& H_{\text{ex}}+H_{\text{DM}}+H_{\text{Z}}+H_{\text{K}},
\end{align}
where
\begin{align}
    H_{\text{ex}} =& -J\sum_{\langle i,j\rangle} \hat{\bdsym{S}}_{i}\cdot \hat{\bdsym{S}}_{j},\notag\\
    H_{\text{DM}} =& -D\sum_{\langle\langle  i,j\rangle\rangle}\nu_{ij}\bdsym{z}\cdot\left(\hat{\bdsym{S}}_{i}\times\hat{\bdsym{S}}_{j}\right),\notag\\
    H_{\text{Z}} =& -g\mu_B \bdsym{B}\cdot\sum_{i} \hat{\bdsym{S}}_{i},\notag\\
    H_{\text{K}} =&-\sum_{\langle i,j\rangle \in \alpha\beta(\gamma) } \Big[
    K^{\gamma} \hat{S}^{\gamma}_{i}\hat{S}^{\gamma}_{j}
    +\Gamma^{\gamma} (\hat{S}^{\alpha}_{i}\hat{S}^{\beta}_{j}+\hat{S}^{\beta}_{i}\hat{S}^{\alpha}_{j})\notag\\
    &+\Gamma'^{\gamma} (\hat{S}^{\gamma}_{i}(\hat{S}^{\alpha}_{j}+\hat{S}^{\beta}_{j})+(\hat{S}^{\alpha}_{i}+\hat{S}^{\beta}_{i})\hat{S}^{\gamma}_{j})
    \Big].
\end{align}
Here, the terms are the ferromagnetic Heisenberg interaction $H_{\text{ex}}$ with $J(>0)$, the next-nearest neighborhood DM interaction $H_{\text{DM}}$, the Zeeman interaction $H_{\text{Z}}$, where the sign of $\nu_{ij}=-\nu_{ji}=\pm1$ is defined in Fig.~\ref{fig1} (b), $\bdsym{z}$ is the unit vector along $z$-axis which constitutes the system coordinate $\{\bdsym{x},\bdsym{y},\bdsym{z}\}$, $g$ is the g-factor, and $\mu_B$ is the Bohr magneton.
In addition, the Kitaev interaction $H_{\text{K}}$ has the bond-dependent spin coupling of $a$, $b$, $c$ types indicated by choosing $\gamma$ in the summation indices $\alpha\beta(\gamma)$ equals to $a$, $b$, $c$, respectively.
Here, the $(\alpha,\beta,\gamma)$ denotes a cyclically permuted set of $\{a,b,c\}$ and $a$, $b$, $c$ denote the Kitaev interaction with the spin operator along the Kitaev vectors $\bdsym{a}$, $\bdsym{b}$, $\bdsym{c}$, i.e., $\hat{S}^{\lambda=x}=\bdsym{x}\cdot\hat{\bdsym{S}}$ for $x=a,b,c$ and $\lambda=\alpha,\beta,\gamma$. 
Geometrical information of the model is that $a_0$ is the lattice constant, the position vectors using $x,y,z$-coordinate system from the sublattice $A$ to the nearest neighborhoods are $\bdsym{d}_1=(a_0/2,a_0/(2\sqrt{3}),0)^{T}$, $\bdsym{d}_2=(-a_0/2,a_0/2\sqrt{3},0)^{T}$, and $\bdsym{d}_3=(0,-a_0/\sqrt{3},0)^{T}$, and to the next-nearest neighborhoods are $\bdsym{R}_1=(a_0/2,-a_0\sqrt{3}/2,0)^{T}$, $\bdsym{R}_2=(a_0/2,a_0\sqrt{3}/2,0)^{T}$, and $\bdsym{R}_3=(-a_0,0,0)^{T}$ in Fig.~\ref{fig1} (c).

We consider the spin wave excited from the fully spin-polarized phase in the presence of the external magnetic field $\bdsym{B}$ along the direction denoted by a unit vector $\bdsym{\chi}$, i.e., $\bdsym{B}=B\bdsym{\chi}=B(\sin{\theta}\cos{\phi},\sin{\theta}\sin{\phi},\cos{\theta})$.
We assume that spins are on the uniformly polarized phase along $\bdsym{n}=(\sin{\vartheta}\cos{\varphi},\sin{\vartheta}\sin{\varphi},\cos{\vartheta})$, where $\varphi$ and $\vartheta$ satisfies the condition given in Appendix~\ref{Appendix A}.
To describe the magnon using the bosonic operators, we apply Holstein-Primakoff transform~\cite{Holstein1940PhysRev.58.1098} to spins in the spin-polarized orthogonal coordinate $\{\bdsym{\xi},\bdsym{\zeta},\bdsym{n}\}$, where the unit vectors $\bdsym{\xi},\bdsym{\zeta},\bdsym{n}$ satisfy the relation $\bdsym{n}=\bdsym{\xi}\times\bdsym{\zeta}$,
\begin{equation}\label{HPtransf}
    \begin{bmatrix}
         \hat{S}^{\xi}_i\\ \hat{S}^{\zeta}_i\\ \hat{S}^{n}_i
    \end{bmatrix}
    \approx
    \begin{bmatrix}
        \sqrt{S/2}(\hat{a}^{\dagger}_i+\hat{a}_i)\\ i\sqrt{S/2}(\hat{a}^{\dagger}_i-\hat{a}_i)\\S-\hat{a}^{\dagger}_i \hat{a}_i
    \end{bmatrix}.
\end{equation}
Here $\hat{a}$ and $i$ are the bosonic operator and index of the $A$-sublattice, and the bosonic operator for the $B$-sublattice is $\hat{b}$ replacing with $\hat{a}$.
The Hamiltonian in Eq.~\eqref{H} is rewritten by using Eq.~\eqref{HPtransf} and the rotation matrix $R(\theta,\phi)$~\cite{Zhang2021PhysRevB.103.174402} transforming the basis of the spin from spin-polarized coordinate to the system coordinate, $( \hat{S}^{x}_i, \hat{S}^{y}_i, \hat{S}^{z}_i)^{T}=R(\vartheta,\varphi)( \hat{S}^{\xi}_i, \hat{S}^{\zeta}_i, \hat{S}^{n}_i)^{T}$, where
\begin{equation}
    R(\vartheta,\varphi)=
    \begin{bmatrix}
        \cos{\vartheta}\cos{\varphi} & -\sin{\varphi} & \sin{\vartheta}\cos{\varphi}\\
        \cos{\vartheta}\sin{\varphi} & \cos{\varphi} & \sin{\vartheta}\sin{\varphi}\\
        -\sin{\vartheta} & 0 & \cos{\vartheta}
    \end{bmatrix}.
\end{equation}
Within the linear spin-wave regime, we take bosonic operators up to the quadratic order as shown in Appendix~\ref{Appendix A}.
The proper polarized direction is when $\varphi$ and $\vartheta$ are chosen to remove the linear order of the bosonic creation and annihilation operator in the Hamiltonian operator.
In our model, they are set to remove the linear order of the bosonic operators that emerged in the Zeeman and Kitaev interactions.
See details in Appendix~\ref{Appendix A}.

To simplify the model, we set Kitaev vectors are $\bdsym{a}=(-1/\sqrt{6},1/\sqrt{2},1/\sqrt{3})^{T}$, $\bdsym{b}=(-1/\sqrt{6},-1/\sqrt{2},1/\sqrt{3})^{T}$, and $\bdsym{c}=(\sqrt{2/3},0,1/\sqrt{3})^{T}$ identical to the $\mathrm{CrI}_{3}$~\cite{Aguilera2020PhysRevB.102.024409, Zhang2021PhysRevB.103.134414}, and the external magnetic along $\bdsym{z}$, namely $[\text{111}]$ in  $\mathrm{CrI}_{3}$.
All spins polarize along [111] as a ground state when B is parallel to [111].
In this case, it is sufficient to consider $J$-$K$-$B$ model, neglecting $\Gamma$ and $\Gamma'$ terms in the $J$-$K$-$\Gamma$-$\Gamma'$-$B$ model, because there exists a mapping from $J$-$K$-$B$ model to $J$-$K$-$\Gamma$-$\Gamma'$-$B$ model~\cite{McClarty2018PhysRevB.98.060404} given by $J\rightarrow J-\Gamma$, $K\rightarrow K+2(\Gamma-\Gamma')$, $g\mu_B B\rightarrow g\mu_B B+3\Gamma S+6\Gamma'S$.
Hence, we deal with the $J$-$DM$-$K$-$B$ model, considering DM interaction on the $J$-$K$-$B$ model, with $S=3/2$, $g=2$, $J>0$, $K\geq 0$, $g\mu_B B >0$, $D\in \mathbb{R}$.

\subsection{Magnon band and magnon Berry curvature}
To obtain the magnon band and the eigenstates of the Hamiltonian, we perform the Fourier transform on the bosonic operators for the diagonalization in the reciprocal space, $\hat{a}^\dagger_i=\frac{1}{\sqrt{N}}\sum_{\bdsym{k}}^{B.Z.}e^{-i\bdsym{R}_i\cdot \bdsym{k}}\hat{a}^\dagger_{\bdsym{k}}$, $\hat{b}^\dagger_j=\frac{1}{\sqrt{N}}\sum_{\bdsym{k}}^{B.Z.}e^{-i\bdsym{R}_j\cdot \bdsym{k}}\hat{b}^\dagger_{\bdsym{k}}$, where $N$ is the number of unit cells, and the Fourier transform relation for the annihilation operator is the hermitian conjugate of the creation operator.
The Hamiltonian in the reciprocal space is given by
\begin{equation}
    \hat{H} = \frac{1}{2}\sum_{\bdsym{k}}\hat{\Psi}_{\bdsym{k}}^{\dagger}H_{\bdsym{k}}\hat{\Psi}_{\bdsym{k}}
\end{equation}
up to the constant with respect to $\bdsym{k}$, where $\hat{\Psi}_{\bdsym{k}}=(\hat{a}_{\bdsym{k}},\hat{b}_{\bdsym{k}},\hat{a}^{\dagger}_{-\bdsym{k}},\hat{b}^{\dagger}_{-\bdsym{k}})$, and the Hamiltonian matrix $H_{\bdsym{k}}$ in the reciprocal space is explained in Appendix~\ref{Appendix B}. The eigenpairs of the magnonic Hamiltonian are obtained by the Bogoliubov transformation which conserves the commutation relation of bosonic operators.
We take the diagonalization of the total Hamiltonian using the Cholesky decomposition~\cite{Colpa1978PhysicaA.93.327,Shindou2013PhysRevB.87.174427} in Appendix~\ref{Appendix B}.
We denotes $\bar{\epsilon}_{n}(\bdsym{k})=\bar{E}_{nn}(\bdsym{k})=\eta_{nn}E_{nn}(\bdsym{k})$ and $|u_n(\bdsym{k})\rangle$ as the $n$th band eigenvalue and right eigenvector of $\bar{H}_{\bdsym{k}} =\eta \mathcal{H}_{\bdsym{k}}$, where $E_{nn}(\bdsym{k})$ is the $n$th band magnon energy , $\eta=\text{diag}(1,1,-1,-1)$, $\langle u_n(\bdsym{k})|\mathcal{H}_{\bdsym{k}}|u_m(\bdsym{k})\rangle=E_{nn}(\bdsym{k})\delta_{nm}$, and $\langle u_n(\bdsym{k})|=(|u_n(\bdsym{k})\rangle)^{\dagger}$, and the detailed calculations are on Appendix~\ref{Appendix B}.

For the $J$-$DM$-$B$ honeycomb ferromagnet model, it is known that nonzero DM interaction opens the magnon gap and the magnon band is topologically non-trivial~\cite{Kim2016PhysRevLett.117.227201,Kim2022PhysRevB.106.104430,Zhu2023EurPhysJPlus.138.1045,Ruckriegel2018PhysRevB.97.081106} with a topological phase transition at $D=0$.
Likewise in a $J$-$K$-$B$ honeycomb ferromagnet model, nonzero Kitaev interaction opens the topological band gap but no phase transition exists within $J>0$ and $K>0$~\cite{McClarty2018PhysRevB.98.060404,Joshi2018PhysRevB.98.060405}.
The $J$-$DM$-$K$-$B$ model also has the topological phase transition line in the half plane consisting of $D\in\mathbb{R}$ and $K\geq 0$, which is $3\sqrt{3} DS =-(KS/2)^2/(3JS+KS+g\mu_B B)$~\cite{Zhang2021PhysRevB.103.134414}.
The gap closes at the $K$ and $K'$ points in the Brillouin zone. The topological property of the magnon band is evaluated through the Berry curvature and the Chern number.
The Berry curvature is defined by the curl of the Berry connection in the reciprocal space, i.e., $n$th band Berry curvature is defined by $\Omega_n(\bdsym{k})=\partial_{\bdsym{k}}\times\bdsym{A}_n(\bdsym{k})$, where the $n$th band Berry connection $\bdsym{A}_n(\bdsym{k})$ is defined by $\bdsym{A}_n(\bdsym{k})=\eta_{nn}\langle u_n(\bdsym{k})|\eta|i\partial_{\bdsym{k}} u_n(\bdsym{k})\rangle$~\cite{Shindou2013PhysRevB.87.174427,Go2024NanoLett.24.5968} in Appendix~\ref{Appendix C}.
The Chern number is defined by $C_n = \iint \frac{d\bdsym{k}}{2\pi} \Omega_n(\bdsym{k})$.
The insertion of $\eta$ in the Berry connection is from $\eta_{nn}\langle u_n(\bdsym{k})|\eta|u_m(\bdsym{k})\rangle=\delta_{nm}$ which means that $\eta_{nn}\langle u_n(\bdsym{k})|\eta$ can be used to capture deviation of the state from the $|u_n(\bdsym{k})\rangle$.
The Chern number of lower energy band changes when the parameters across the topological phase transition line.
For convenience, we call the lower energy band and upper energy band the lower band and upper band respectively.

\subsection{Magnon orbital moment Berry curvature}

The magnon orbital moment Berry curvature (MOMBC) appears in the linear response of magnon orbital moment current density $\bdsym{J}^{\bdsym{m}}$ induced by external potential, which is, in our case, temperature gradient as in Fig.~\ref{fig1} (a). In particular, in this work, we consider the $y$ directional magnon current density carrying the $z$ component of the magnon orbital moment $\hat{\bdsym{m}}=(\hat{\bdsym{r}}\times \hat{\bdsym{v}}-\hat{\bdsym{v}}\times \hat{\bdsym{r}})/4$~\cite{Sahu2021PhysRevB.103.085113,Pezo2022PhysRevB.106.104414}, namely $J_y^{m_z}$, when the temperature gradient is along $x$ direction as described in Fig.~\ref{fig1} (a).

\begin{figure*}
    \centering
    \includegraphics[width=510pt]{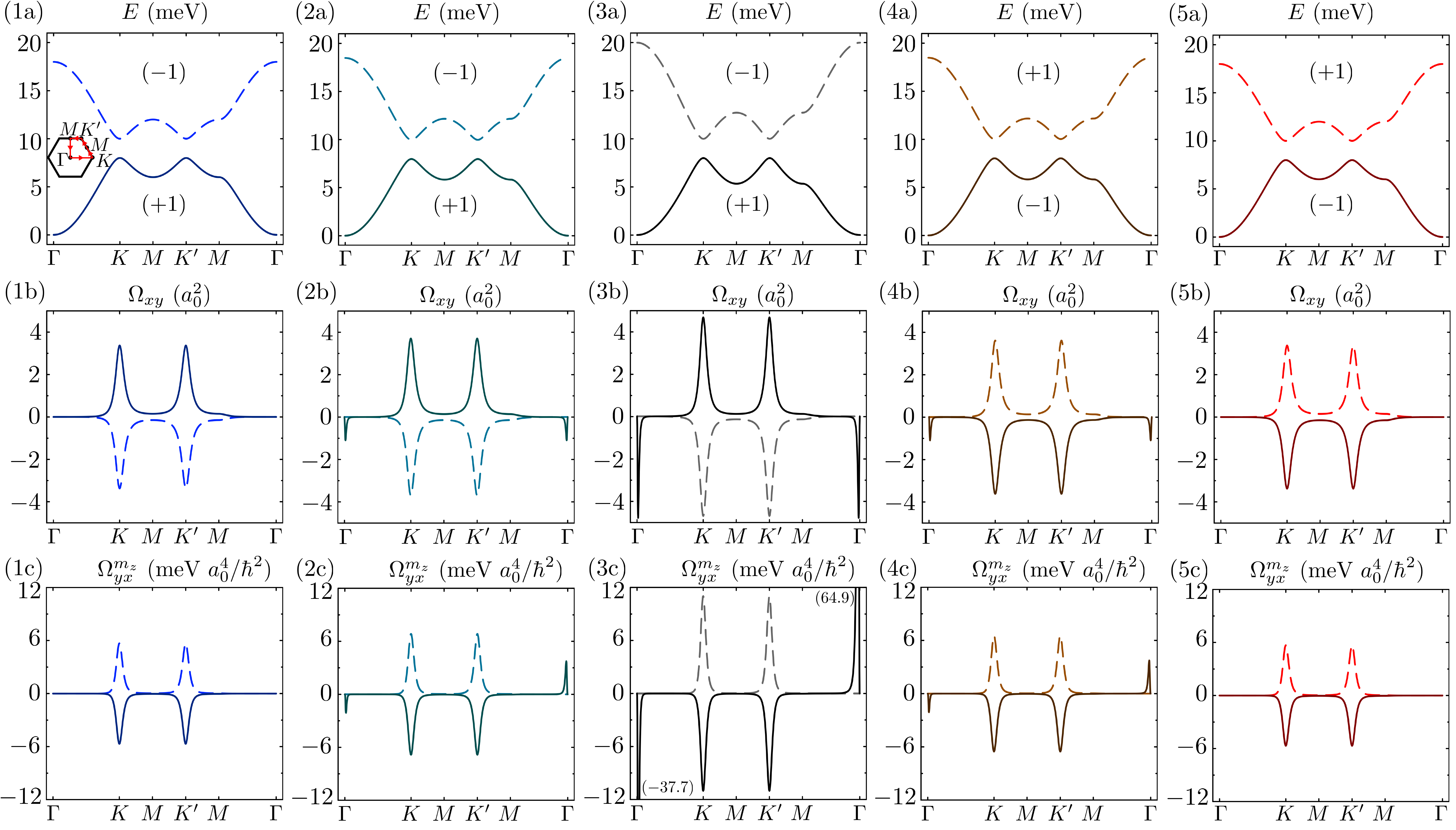}
    \caption{
    The numbers from 1 to 5 denote the system with parameters $(J,D,K)=$ $(1.999,0.1283,0)$, $(1.387,0.09293,2)$, $(0.8876,0.0,4)$, $(1.387,-0.1568,2)$, $(1.999,-0.1283,0)$ in meV, respectively, when the magnetic field $B=0.05$ T is applied.
    The numbers are identical to the ones used in Fig.~\ref{General models MOMNC}.
    For each row, energy band (1-5a), Berry curvature (1-5b), and MOMBC (1-5c) are plotted.
    The parentheses in (1-5a) denote the Chern numbers of the corresponding bands.
    The numbers in the parentheses in (3c) denote the value at the lower band's near $\Gamma$ point peaks.
    They are plotted along the symmetric points in the reciprocal space connected by a directed red line on the hexagon in (1a), which starts and ends at $\Gamma$ point $(k_x,k_y)=(0,0)$ passing $K$ point $(4\pi/3a_0,0)$, $M$ point $(\pi/a_0,\pi/\sqrt{3}a_0)$, $K'$ point $(2\pi/3a_0,2\pi/\sqrt{3}a_0)$, and $M$ point $(0,2\pi/\sqrt{3}a_0)$ sequentially.
    To indicate the bands, we used darker solid lines for the lower band and brighter dashed lines for the upper band.
    }
    \label{General two model structure}
\end{figure*}

We treat the temperature gradient as fictitious pseudo-gravitational potential~\cite{Go2024NanoLett.24.5968,Luttinger1964PhysRev.135.A1505, Matsumoto2014PhysRevB.89.054420, Zyuzin2016PhysRevLett.117.217203,Li2020PhysRevRes.2.013079}.
The thermal response of magnon orbital moment current density $J^{m_z}_{y}$ from the temperature gradient along $x$-direction that we consider is calculated in linear response regime by $J^{m_z}_{y} = -\alpha^{m_z}_{yx}\partial_{x}T$~\cite{Go2024NanoLett.24.5968,Zyuzin2016PhysRevLett.117.217203,Li2020PhysRevRes.2.013079}, where $\alpha^{m_z}_{yx}$ is the MOMNC defined by
\begin{equation}\label{Nernstcond}
    \alpha^{m_z}_{yx}=\frac{2k_B}{V}\sum_{n}\sum_{\bdsym{k}}c_1(g(\epsilon_n(\bdsym{k}),T_0))\Omega^{m_z}_{n}(\bdsym{k}),
\end{equation}
where $k_B$ is the Boltzmann constant, $\hbar$ is the reduced Plank constant, $V$ is the area of the system which is considered as a 2-dimension,
$c_1(g)=(1+g)\ln (1+g)-g\ln g$, $g(\epsilon,T)=(e^{\epsilon/k_B T}-1)^{-1}$ is the Bose-Einstein distribution function, $\epsilon_n(\bdsym{k})$ is $E_{nn}(\bdsym{k})$, $T_0$ is a mean temperature, i.e., $T=T_0+\delta T(x)$, where small $\delta T(x)$, i.e., $|\delta T(x)|\ll T_0$, carries the spatial dependency of the temperature causing the temperature gradient on the system, and $\Omega^{m_z}_n(\bdsym{k})$ is MOMBC of $n$th band for $m_z$. MOMBC of $n$th band for $m_z$ is given by Ref.~\cite{Li2020PhysRevRes.2.013079,Sahu2021PhysRevB.103.085113},
\begin{equation}\label{Lberry}
\begin{split}
    &\Omega^{m_z}_n(\bdsym{k})=-2\hbar \sum_{m\neq n}\text{Im}\Big[\frac{\eta_{nn}\eta_{mm}}{(\bar{\epsilon}_{n}(\bdsym{k})-\bar{\epsilon}_{m}(\bdsym{k}))^2}\\
    &\times\langle u_n(\bdsym{k})|\hat{j}^{m_z}_{y}|u_m(\bdsym{k})\rangle\langle u_m(\bdsym{k})|\hat{v}_x|u_n(\bdsym{k})\rangle\Big],
\end{split}
\end{equation}
where $\hat{j}^{m_z}_{y}$ is a component of a corresponding case in the magnon orbital moment current operator defined by $\hat{\bdsym{j}}^{\bdsym{m}}=(\hat{\bdsym{m}}\eta \hat{\bdsym{v}}+\hat{\bdsym{v}}\eta \hat{\bdsym{m}})/4$.
The explicit evaluation is in Appendix~\ref{Appendix C}.

We evaluate the magnon energy band (a), Berry curvature (b), and MOMBC (c) for in Fig.~\ref{General two model structure} the following five cases with $B=0.05$ T: $(J,D,K)=$ $(1.999,0.1283,0)$, $(1.387,0.09293,2)$, $(0.8876,0,4)$, $(1.387,-0.1568,2)$, $(1.999,-0.1283,0)$ in meV which we call (1), (2), (3), (4), and (5), respectively.
The two cases (1) and (5) with no Kitaev interaction belong to the $J$-$DM$-$B$ model.
The (3) case with no DM interaction corresponds to the $J$-$K$-$B$ model.
The other two cases (2) and (4) belong to the $J$-$DM$-$K$-$B$ model.
The parameters $J$, $DM$, and $K$ are set to have the same lower bandwidth and band gap.
The figures in Fig.~\ref{General two model structure} are the magnon energy band (a), Berry curvature (b), and MOMBC (c) along a directed red line on the hexagon in Fig.~\ref{General two model structure} (1a), which starts and ends at $\Gamma$ point $(k_x,k_y)=(0,0)$ passing $K$ point $(4\pi/3a_0,0)$, $M$ point $(\pi/a_0,\pi/\sqrt{3}a_0)$, $K'$ point $(2\pi/3a_0,2\pi/\sqrt{3}a_0)$, and $M$ point $(0,2\pi/\sqrt{3}a_0)$ sequentially.

\subsubsection{$J$-$DM$-$B$ model}
The $J$-$DM$-$B$ models are (1a-c) with $D>0$ and (5a-c) with $D<0$ in Fig.~\ref{General two model structure}.
The Chern number of the lower band follows the sign of DM interaction $D$.
Two cases have the same band structures and MOMBC but have the opposite signed Berry curvatures.
The Berry curvatures and MOMBCs are large for both models at $K$ and $K'$ points, and zero for $\Gamma$ point.

Both models' magnon orbital moment Nernst currents flow in the same direction for a given temperature gradient.
In contrast, magnon thermal Hall currents of the models, for instance, flow in different directions.
A magnon orbital moment Nernst current consists of two magnon orbital moment currents which are a magnon orbital moment current and an oppositely signed magnon orbital moment flowing along the opposite direction in Fig.~\ref{fig1} (a).
In this sense, the magnon orbital moments of the two models have different signs.

\subsubsection{$J$-$K$-$B$ model}
The $J$-$K$-$B$ model is (3a-c) in Fig.~\ref{General two model structure}.
We observe the peaks near the $\Gamma$ point of the Berry curvature and MOMBC in the lower band.
Unlike the Berry curvature, the near $\Gamma$ point peaks in the MOMBC have sign variation, which is positive along $k_y$-direction with maximum value $64.9$ $\text{meV }a_0^4/\hbar^2$ and negative along $k_x$-direction with minimum value $-37.7$ $\text{meV }a_0^4/\hbar^2$ from the origin in Fig.~\ref{General two model structure} (3c).
With the same interpretation used in the $J$-$DM$-$B$ model, the near $\Gamma$ point peaks with different signs of MOMBC show that the magnon residing in each peak has opposite signs of the magnon orbital moment.

In contrast to the $J$-$DM$-$B$ model where the magnetic field shifts the overall energy~\cite{Kim2016PhysRevLett.117.227201, Owerre2016JApplPhys.120.043903}, the magnetic field in the $J$-$K$-$B$ model inhomogeneously raises the magnon energy in the reciprocal space.
Thus the Berry curvature and MOMBC change when the strength of the magnetic field does.
The prominent change is the drastic suppression of the near $\Gamma$ point peaks by the increase of the magnetic field.
The details are written in Appendix~\ref{Appendix F}.

\subsubsection{$J$-$DM$-$K$-$B$ model}
The $J$-$DM$-$K$-$B$ models are (2a-c) with $D>0$ and (4a-c) with $D<0$ in Fig.~\ref{General two model structure}.
These models have both characteristics of $J$-$DM$-$B$ model and $J$-$K$-$B$ model.
When changing the sign of the $D$ alters the topological phase of the system, i.e., $3\sqrt{3} |D|S>(KS/2)^2/(3JS+KS+g\mu_B B)$, the signs of the Berry curvature at $K$ and $K'$ points also changes, while the MOMBC does not.
At the same time, like the $J$-$K$-$B$ model, the $J$-$DM$-$K$-$B$ models have the near $\Gamma$ point peaks on the lower band in Fig.~\ref{General two model structure} (b2-3, c2-3), of which sign does not depend on the sign of the DM interaction. 

The qualitative shapes, namely signs of peaks depending on the directions in the reciprocal space, of the near $\Gamma$ point peaks are similar to the $J$-$K$-$B$ model due to the common origin of the peak, the Kitaev interaction.

\subsection{Magnon orbital moment Nernst Conductivity}
\begin{figure}
    \centering
    \includegraphics[width=240pt]{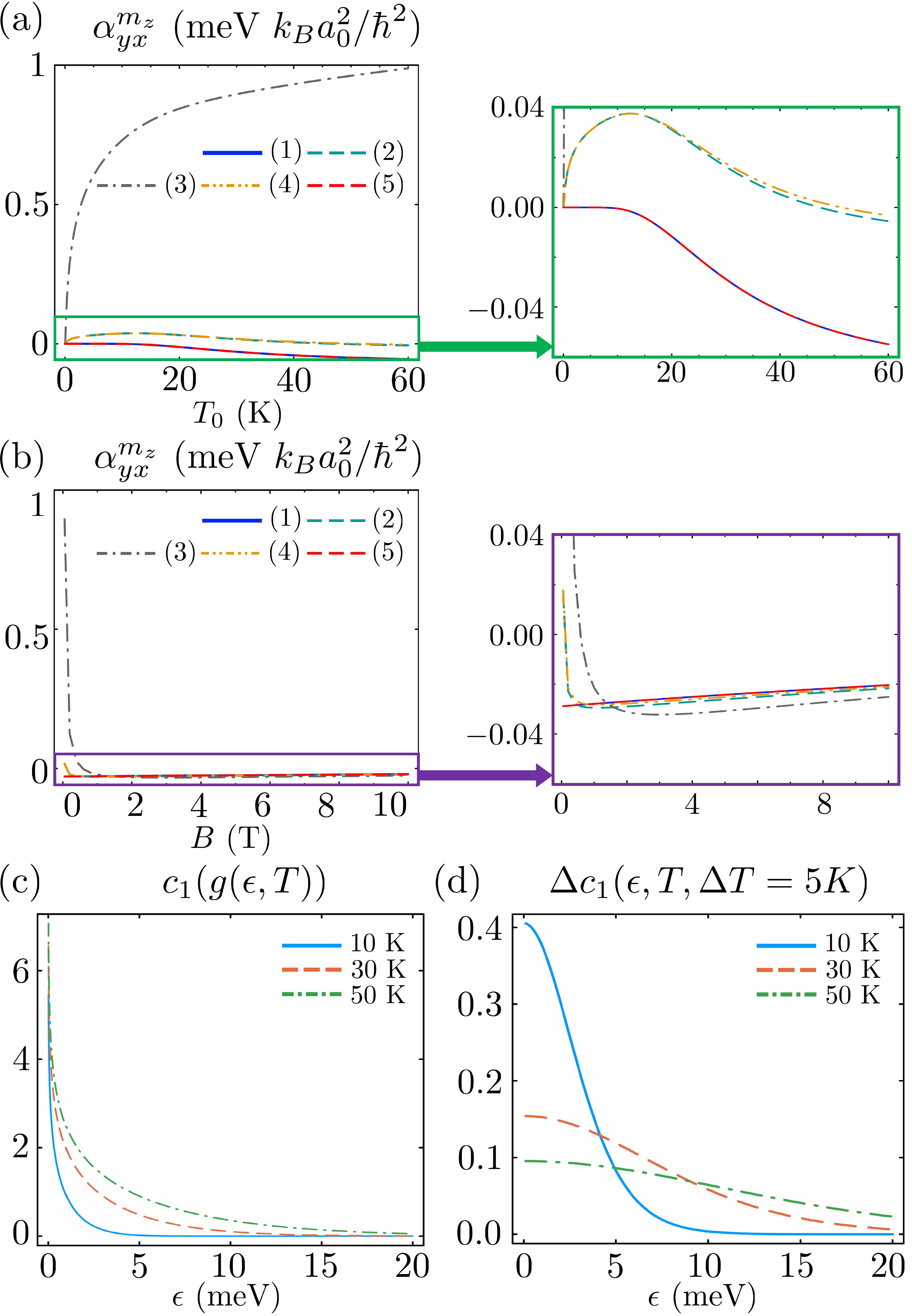} 
    \caption{The MOMNC [Eq.~\eqref{Nernstcond}] dependence on the mean temperature with fixed magnetic field $B=0.05$ T (a), and on the magnetic field with fixed mean temperature $T_0=30$ K (b).
    The parameters $(J,D,K)=$ $(1.999,0.1283,0)$, $(1.387,0.09293,2)$, $(0.8876,0,4)$, $(1.387,-0.1568,2)$, $(1.999,-0.1283,0)$ in meV for (1-5) respectively.
    The parameters (1-5) are identical to that used in Fig.~\ref{General two model structure}.
    The zoomed-in plots share the line color and shape, the axes, and the units.
    Figure (c) $c_1(g(\epsilon,T))$ and (d) their difference $\Delta c_1(\epsilon,T,\Delta T)$ from the temperature $T$ to $T+\Delta T$ for $T=$ 10 K, 30 K, 50 K, which are denoted by sky blue solid line, orange dashed line, and green dash-dotted line respectively.
    }
    \label{General models MOMNC}
\end{figure}

In Fig.~\ref{General models MOMNC} (a, b), we evaluate the MOMNC for varying the mean temperature and the strength of the magnetic field, respectively, with the following five cases (1-5) used in Fig.~\ref{General two model structure}: $(J,D,K)=$ $(1.999,0.1283,0)$, $(1.387,0.09293,2)$, $(0.8876,0,4)$, $(1.387,-0.1568,2)$, $(1.999,-0.1283,0)$ in meV which we call (1), (2), (3), (4), and (5), respectively.
First, note that the MOMNCs of two $J$-$DM$-$B$ models (1, 5) are identical in Fig.~\ref{General models MOMNC}.
Second, the MOMNCs of two $J$-$DM$-$K$-$B$ models (2, 4) seem to be overlapped for weak magnetic fields and low mean temperatures, but $J$-$DM$-$K$-$B$ model (4) has larger MOMNC than $J$-$DM$-$K$-$B$ model (2) for other regimes in Fig.~\ref{General models MOMNC} due to the difference in their magnon bands and MOMBCs.

We analyze the temperature and magnetic field dependence of the MOMNC using the magnon population contribution $c_1(g(\epsilon,T))$ used in Eq.~\eqref{Nernstcond}, whose energy dependence is shown in Fig.~\ref{General models MOMNC} (c), and its temperature difference $\Delta c_1(\epsilon,T,\Delta T)=c_1(g(\epsilon,T+\Delta T))-c_1(g(\epsilon,T))$ shown in Fig.~\ref{General models MOMNC} (d) when $\Delta T=5$ K, in addition to considering the MOMBC in Fig.~\ref{General two model structure}.

\subsubsection{Temperature dependence of MOMNC}
The dependence of the MOMNC on mean temperature $T_0$ is shown in Fig.~\ref{General models MOMNC}(a).
We analyze the temperature dependence of MOMNC through the magnon population contribution $c_1(g(\epsilon,T))$.
The magnon population contribution $c_1(g(\epsilon,T))$ with respect to the energy is a positive definite decreasing function having a sharp peak at zero energy $\epsilon=0$ and being flat for higher energy in Fig.~\ref{General models MOMNC} (c).
As $T$ goes to $0$, $c_1(g(\epsilon,T))$ converges to $0$ except $\epsilon=0$.
This makes MOMNCs zero at $T_0=0$ K for all models we considered due to no magnon population contribution because the $\epsilon>0$ for $g\mu_B B>0$, for comparison, electronic systems which need a delicate consideration in the limit of $T_0=0$ because spin Hall conductivity calculation in bare bubble approximation with vertex correction diverges but converges to zero with considering orbital correction~\cite{Dyrdal2016PhysRevB.94.035306, Dyrdal2016PhysRevB.94.205302}.

The change of the MOMNC is understood by the additional MOMNC by the increase of the magnon population contribution for a temperature increase, i.e., the difference of the magnon population contribution $\Delta c_1(\epsilon,T,\Delta T)$.
The profile of $\Delta c_1(\epsilon,T,\Delta T)$ is a positive definite, decreasing function in Fig.~\ref{General models MOMNC} (d).
The $\Delta c_1(\epsilon,T,\Delta T)$ for a fixed temperature increase $\Delta T$ becomes flatter as the $T$ increases.
This indicates that the MOMNC contribution of the lower energy magnon relative to that of the higher energy magnon decreases as the temperature $T$ increases.

For the zero Kitaev interaction cases, i.e., (1, 5) cases in Fig.~\ref{General models MOMNC} (a), the MOMNCs decrease as the temperature increases without sign change because the lower band MOMBC is negative semi-definite.
Additionally, their MOMBCs have no near $\Gamma$ point peaks, and have significant value near the $K$ and $K'$.
Thus, MOMNC decreases small amount for the low temperature.
As the mean temperature increases in Fig.~\ref{General models MOMNC} (a), the magnon population near the $K$ and $K'$ points increases, which decreases the MOMNCs, meaning that the magnon orbital moment Nernst current along $-y$-direction increases.
Due to the flattening behavior of $\Delta c_1(\epsilon,T,\Delta T)$ for a fixed $\Delta T$, the MOMNC decrease slows down for high temperatures.

For nonzero Kitaev interaction $K\neq 0$ cases, i.e., (2-4) cases in Fig.~\ref{General models MOMNC} (a), a change of the contribution of the near $\Gamma$ point relative to that of the $K$ and $K'$ points explains the MOMNC behavior as follows.
At low temperatures $T>0$, the MOMNC is mainly affected by near $\Gamma$ point peaks because the magnon population on the lower band near $\Gamma$ points is much larger than that on the $K$ and $K'$ points.
For a low $T_0$ region, the MOMNC increases as the mean temperature $T_0$ increases, because the magnon population contribution increase with respect to the energy is steep enough for the MOMNC contribution near $\Gamma$ point peaks to be larger than that of the $K$ and $K'$ points.
As the $T_0$ keeps increasing further, however, the MOMNC for the $J$-$DM$-$K$-$B$ models begins to decrease from a certain mean temperature, at which the increase of the contribution for the MOMNC from near $K$ and $K'$ points becomes comparable to that from the near $\Gamma$ point peaks due to the flattening behavior of $\Delta c_1(\epsilon,T,\Delta T)$.
For the $J$-$K$-$B$ model, case (3) in Fig.~\ref{General models MOMNC} (a), due to the large near $\Gamma$ point peak, the MOMNC seems to keep increasing within our temperature range, i.e., below the Curie temperature $\approx 61$ K~\cite{McGuire2015ChemMater.27.612}. This unusual temperature dependence of the MOMNC of the $J$-$K$-$B$ model can be used as an experimental probe for the significant Kitaev interaction and the resultant peaks in the MOMBC near the $\Gamma$-point.

\subsubsection{Magnetic field strength dependence of MOMNC}
The external magnetic field $B$ affects the MOMNC as shown in Fig.~\ref{General models MOMNC} (b) through the magnon population and the MOMBC.
The magnon energy band raised by the external magnetic field induces magnon population change.
For a given energy increase, the population contribution decrease of lower energy is larger than that of a higher energy, i.e., $dc_1(g(\epsilon,T))/d\epsilon|_{\epsilon_1}<dc_1(g(\epsilon,T))/d\epsilon|_{\epsilon_2}<0$ for $\epsilon_1<\epsilon_2$, due to the shape of the magnon population contribution $c_1(g(\epsilon))$ in Fig.~\ref{General models MOMNC} (c).

For the zero Kitaev interaction cases, i.e., (1, 5) cases in Fig.~\ref{General models MOMNC} (b), the MOMNCs increase steadily as the magnetic field increases.
The reason is the decrease of the magnon population by the constant magnon energy increased through the Zeeman interaction $g \mu_B B$~\cite{Owerre2016JApplPhys.120.043903, Kim2016PhysRevLett.117.227201}, and the population contribution decrease is larger for the lower band than the upper band.

For nonzero Kitaev interaction, $K\neq 0$, cases, the MOMNC (2-4) in Fig.~\ref{General models MOMNC} (b) is understood in two mechanisms; one is a shortening of the near $\Gamma$ point peaks, and the other is the population contribution change.
The shortening of the near $\Gamma$ point peaks of MOMBCs as shown in Fig.~\ref{figF2} causes the drastic decline of the MOMNCs within a weak magnetic field region for the cases (2-4) in Fig.~\ref{General models MOMNC}, similar to thermal Hall effect~\cite{McClarty2018PhysRevB.98.060404}.
For higher magnetic field regions, where the near $\Gamma$ point peaks are negligibly small, the behavior of the MOMNCs is mainly due to the magnon population contribution change.

Another mechanism, the population contribution change, is inhomogeneous in the reciprocal space, unlike the zero Kitaev interaction cases.
The nonzero Kitaev interaction cases have energy increases depending on reciprocal space in addition to the constant Zeeman energy increase $g \mu_B B$.
The additional energy increase relative to the energy is small except for the near $\Gamma$ point on the lower band, where the magnon energy is low, as shown in Fig.~\ref{figF1}.
This escalates the contribution decrease for the near $\Gamma$ point on the lower band in Fig.~\ref{General models MOMNC} (c).
As the magnetic field increases, the additional energy increase relative to the energy becomes homogeneous as shown in Fig.~\ref{figF1}.
In the strong magnetic field, the MOMNCs of the nonzero Kitaev cases increase like that of the zero Kitaev cases.

\section{Application on $\mathrm{CrI}_{3}$}\label{sect 3}
\begin{figure}
    \centering
    \includegraphics[width=240pt]{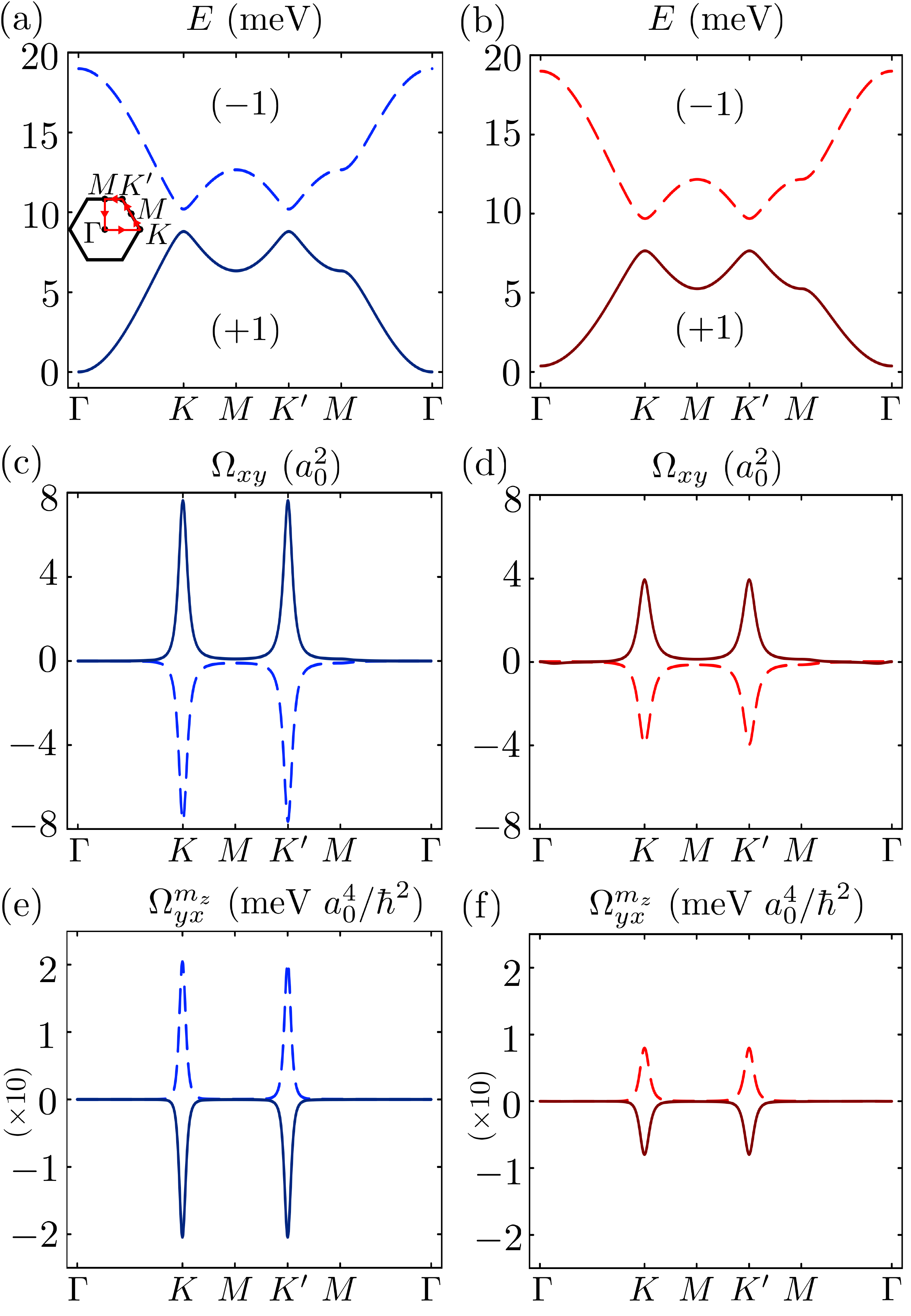}
    \caption{The energy band (a, b), Berry curvature (c, d), MOMBC (e, f) plots of the $J$-$DM$-$B$ model (a, c, e) and $J$-$K$-$\Gamma$-$B$ model (b, d, f) when the magnetic field $B=0.05$ T is applied.
    The parentheses in (a) and (b) denote the Chern numbers of the corresponding bands.
    They are plotted along the symmetric points in the reciprocal space connected by a directed red line on the hexagon in (a), and the path is identical to one in Fig.~\ref{General two model structure}.
    The colors of lines in $J$-$DM$-$B$ model (a, c, e) and $J$-$K$-$\Gamma$-$B$ model (b, d, f) are bluish and reddish respectively, and to indicate band we used darker solid lines for the lower band and brighter dashed line for the upper band.
    The parameter values of the $J$-$DM$-$B$ model and the $J$-$K$-$\Gamma$-$B$ model of $\mathrm{CrI}_{3}$ are used.
    }
    \label{CrI3 structure fig}
\end{figure}

In this section, we estimate the magnon band and the magnon Berry curvature for the $J$-$DM$-$B$ model and the $J$-$K$-$\Gamma$-$B$ model of $\mathrm{CrI}_{3}$~\cite{Chen2021PhysRevX.11.031047}.
From Ref.~\cite{Chen2021PhysRevX.11.031047}, we use the parameter $J=2.11$ meV, $D=0.09$ meV for the $J$-$DM$-$B$ model, and $J=0.83$ meV, $K=3.8$ meV, $\Gamma=0.082$ meV for the $J$-$K$-$\Gamma$-$B$ model, $S=3/2$, and $g=2$ commonly.
We consider the case where the magnetic field is applied along the $z$-axis, i.e., $[\text{111}]$ direction, $\bdsym{B}=B\bdsym{z}$.

\begin{figure}
    \includegraphics[width=240pt]{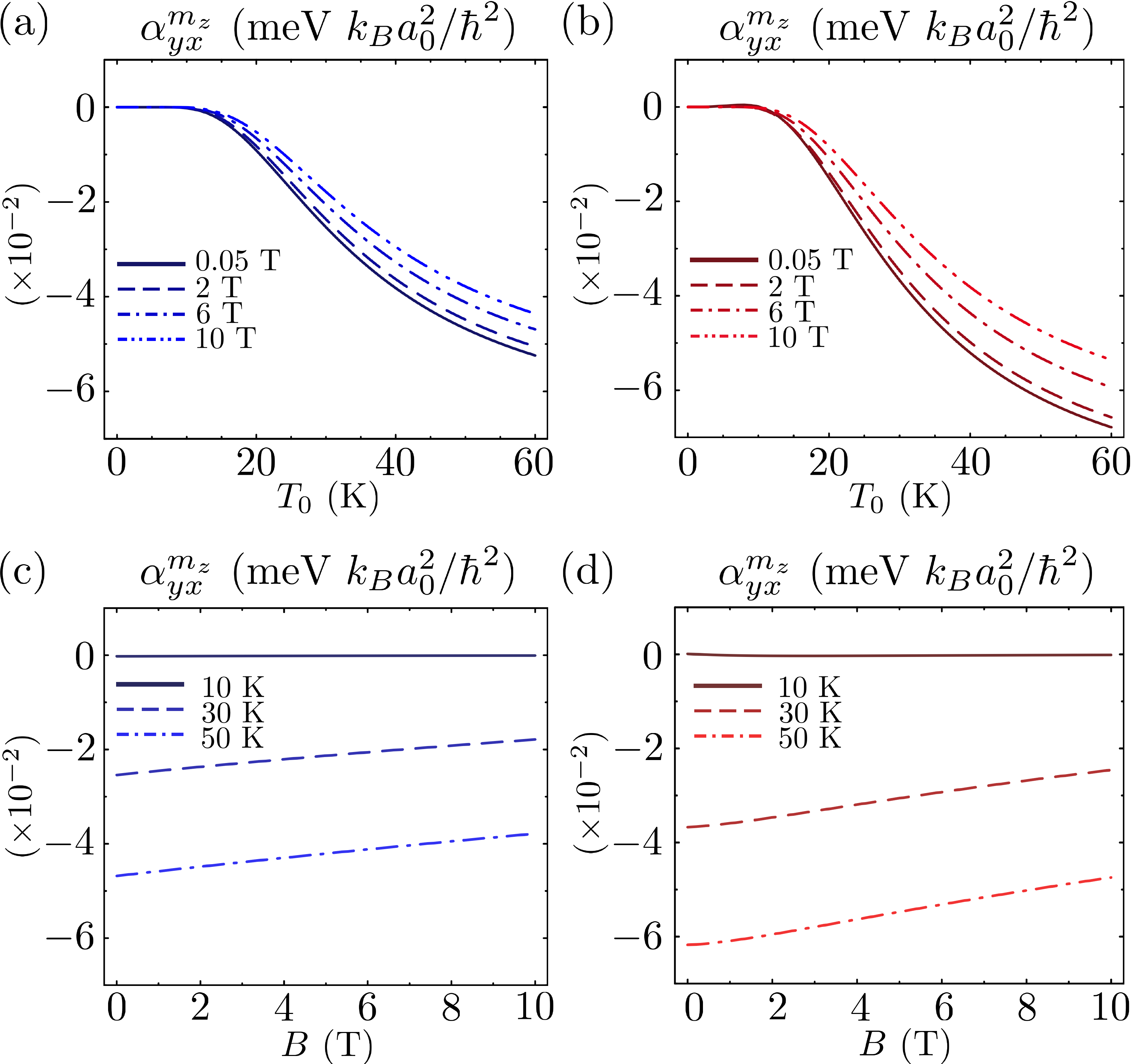}
    \caption{The MOMNCs $\alpha^{m_z}_{yx}$ of $J$-$DM$-$B$ model (a, c) and $J$-$K$-$\Gamma$-$B$ model (b, d), where $k_B$ is the Boltzmann constant.
    In (a, b) we denote the lines solid, dashed, dashed with a single dot, and dashed with double dots for $B=$ $0.05$, $2$, $6$, $10$ T cases, respectively, and the colors of the graph are vivified as the magnetic field increases.
    In (c, d) we denote the lines solid, dashed, and dashed with a single dot for the mean temperature $T_0=$ $10$, $30$, $50$ K cases, respectively, and the colors of the graph are vivified as the temperature increases.}
    \label{CrI3 MOMNC fig}
\end{figure}

In Fig.~\ref{CrI3 structure fig} (a, b), the magnon energy bands for $J$-$DM$-$B$ model and $J$-$K$-$\Gamma$-$B$ model show similar magnon dispersion relations.
The Berry curvatures in Fig.~\ref{CrI3 structure fig} (c, d) show the same Chern numbers, $+1$ for lower bands and $-1$ for upper bands in Fig.~\ref{CrI3 structure fig} (a, b).
It is natural to expect the $J$-$DM$-$B$ model and $J$-$K$-$\Gamma$-$B$ model of $\mathrm{CrI}_{3}$ be similar to the $J$-$DM$-$B$ model and the $J$-$K$-$B$ model in the previous section.
The $J$-$DM$-$B$ model of $\mathrm{CrI}_{3}$ in Fig.~\ref{CrI3 structure fig} (a, c, e) have similar shapes but different magnitudes due to the system parameter difference compared to the $J$-$DM$-$B$ model in Fig.~\ref{General two model structure} (1a-c).
However, the near $\Gamma$ point peaks in the Berry curvature and MOMBC in Fig.~\ref{CrI3 structure fig} (d, f) are hardly visible, and they look similar to the $J$-$DM$-$B$ model rather than the $J$-$K$-$B$ model.
The reasons are that $\Gamma$ in $J$-$K$-$\Gamma$-$B$ model behaves as a magnetic field ($\sim 6$ T) in $J$-$K$-$B$ model, and the near $\Gamma$ point peaks are fragile to the external magnetic field in Fig.~\ref{figF2}. We can understand the MOMNC $\alpha^{m_z}_{yx}$ in Fig.~\ref{CrI3 MOMNC fig} based on that of the $J$-$DM$-$B$ model and the $J$-$K$-$B$ model in the previous section. For the $J$-$K$-$\Gamma$-$B$ model, the near $\Gamma$ point effect is suppressed, so the overall behavior of MOMNC is identical to the $J$-$DM$-$B$ model.

\section{Conclusion}\label{sect 4}
In this paper, we have obtained the magnon orbital moment Nernst conductivity, which quantifies the magnon orbital moment Nernst current, of a honeycomb ferromagnet with Heisenberg, DM, Kitaev, and Zeeman interaction within the linear spin wave from the fully polarized spin ground phase by handling the temperature gradient treated by a pseudo-gravitational potential. One of our main findings is that the honeycomb ferromagnets exhibit the magnon orbital moment Nernst conductivity either via the DM interaction or via the Kitaev interaction.
It is distinct from the honeycomb-antiferromagnet case where the magnon orbital moment Nernst conductivity can be finite without any spin-orbit coupling~\cite{Go2024NanoLett.24.5968}.
The magnon orbital moment Berry curvature of the $J$-$DM$-$B$ models does not depend on the sign of DM interaction in contrast to the Berry curvature.
This indicates the carried magnon orbital moment changes by the sign of DM interaction.
They have the same magnon orbital moment Nernst conductivities which decrease as the magnon population contribution increases, i.e., the mean temperature increases or the magnetic field decreases.

In the $J$-$K$-$B$ model, the near $\Gamma$ point peaks of the Berry curvature have the same sign but the magnon orbital moment Berry curvature does not, indicating that the peaks carry different magnon orbital moments.
In the $J$-$DM$-$K$-$B$ models, which have both characteristics of $J$-$DM$-$B$ model and $J$-$K$-$B$ model, the near $\Gamma$ point peaks of the Berry curvature are independent of the sign of the Chern number.
The magnon orbital moment Berry curvature also shows the independence of the Chern number sign.
The near $\Gamma$ point peaks of $K\neq 0$ models give positive magnon orbital moment Nernst conductivity for low magnetic field and low mean temperature where the $K$ and $K'$ point magnon contributions are small relative to near $\Gamma$ point contributions, and the peaks drop with turning on the magnetic field.

Applying our formalism to two models of $\mathrm{CrI}_{3}$~\cite{Chen2021PhysRevX.11.031047}, $J$-$DM$-$B$ model and $J$-$K$-$\Gamma$-$B$ model, the magnon orbital moment Nernst conductivities are similar functions with different magnitudes.
The reason is that the $\Gamma$ term in $J$-$K$-$\Gamma$-$B$ model acts like the magnetic field in the $J$-$K$-$B$ model, so the near $\Gamma$ point peaks are short for making a notable positive lump in magnon orbital moment Nernst conductivity.

Our research investigated the magnon orbital moment Nernst conductivity related to the electrical polarization induced by magnon orbital moment accumulated on the edges of the system~\cite{Go2024NanoLett.24.5968}.
This work is the stepping stone for understanding the effect of the magnon Hamiltonian components on the electrical interaction, such as photon as commented in Ref.~\cite{Fishman2022PhysRevLett.129.167202, Go2024NanoLett.24.5968, To2024ArXiv.16004}, through the magnon orbital moment.

On the side of the nonzero Kitaev interaction system, our research showed the low $\Gamma$ and $\Gamma'$ Kitaev material to provide tunability through a sensitivity of the near $\Gamma$ point peaks to the magnetic field.
This means that reducing the effects of the $\Gamma$ and $\Gamma'$ terms is the key to distinguishing the models by measuring the dependence of the magnon orbital moment Nernst conductivity on the magnetic field. Lastly, it would be interesting to investigate how magnetic structural defects such as magnetic domain walls, impurities, and surface defects affect the magnon orbital moment Nernst conductivity, which is beyond the scope of the current work.

\section{Acknowledgement}
We thank Gyungchoon Go for helpful discussions.
This work was supported by the Samsung Science and Technology Foundation (SSTF-BA2202-04).

\bibliography{Reference}

\begin{thebibliography}{66}%
\makeatletter
\providecommand \@ifxundefined [1]{%
 \@ifx{#1\undefined}
}%
\providecommand \@ifnum [1]{%
 \ifnum #1\expandafter \@firstoftwo
 \else \expandafter \@secondoftwo
 \fi
}%
\providecommand \@ifx [1]{%
 \ifx #1\expandafter \@firstoftwo
 \else \expandafter \@secondoftwo
 \fi
}%
\providecommand \natexlab [1]{#1}%
\providecommand \enquote  [1]{``#1''}%
\providecommand \bibnamefont  [1]{#1}%
\providecommand \bibfnamefont [1]{#1}%
\providecommand \citenamefont [1]{#1}%
\providecommand \href@noop [0]{\@secondoftwo}%
\providecommand \href [0]{\begingroup \@sanitize@url \@href}%
\providecommand \@href[1]{\@@startlink{#1}\@@href}%
\providecommand \@@href[1]{\endgroup#1\@@endlink}%
\providecommand \@sanitize@url [0]{\catcode `\\12\catcode `\$12\catcode `\&12\catcode `\#12\catcode `\^12\catcode `\_12\catcode `\%12\relax}%
\providecommand \@@startlink[1]{}%
\providecommand \@@endlink[0]{}%
\providecommand \url  [0]{\begingroup\@sanitize@url \@url }%
\providecommand \@url [1]{\endgroup\@href {#1}{\urlprefix }}%
\providecommand \urlprefix  [0]{URL }%
\providecommand \Eprint [0]{\href }%
\providecommand \doibase [0]{https://doi.org/}%
\providecommand \selectlanguage [0]{\@gobble}%
\providecommand \bibinfo  [0]{\@secondoftwo}%
\providecommand \bibfield  [0]{\@secondoftwo}%
\providecommand \translation [1]{[#1]}%
\providecommand \BibitemOpen [0]{}%
\providecommand \bibitemStop [0]{}%
\providecommand \bibitemNoStop [0]{.\EOS\space}%
\providecommand \EOS [0]{\spacefactor3000\relax}%
\providecommand \BibitemShut  [1]{\csname bibitem#1\endcsname}%
\let\auto@bib@innerbib\@empty
\bibitem [{\citenamefont {Wolf}\ \emph {et~al.}(2001)\citenamefont {Wolf}, \citenamefont {Awschalom}, \citenamefont {Buhrman}, \citenamefont {Daughton}, \citenamefont {von Molnár}, \citenamefont {Roukes}, \citenamefont {Chtchelkanova},\ and\ \citenamefont {Treger}}]{Wolf2001Science.294.1488}%
  \BibitemOpen
  \bibfield  {author} {\bibinfo {author} {\bibfnamefont {S.~A.}\ \bibnamefont {Wolf}}, \bibinfo {author} {\bibfnamefont {D.~D.}\ \bibnamefont {Awschalom}}, \bibinfo {author} {\bibfnamefont {R.~A.}\ \bibnamefont {Buhrman}}, \bibinfo {author} {\bibfnamefont {J.~M.}\ \bibnamefont {Daughton}}, \bibinfo {author} {\bibfnamefont {S.}~\bibnamefont {von Molnár}}, \bibinfo {author} {\bibfnamefont {M.~L.}\ \bibnamefont {Roukes}}, \bibinfo {author} {\bibfnamefont {A.~Y.}\ \bibnamefont {Chtchelkanova}},\ and\ \bibinfo {author} {\bibfnamefont {D.~M.}\ \bibnamefont {Treger}},\ }\bibfield  {title} {\bibinfo {title} {Spintronics: A spin-based electronics vision for the future},\ }\href {https://doi.org/10.1126/science.1065389} {\bibfield  {journal} {\bibinfo  {journal} {Science}\ }\textbf {\bibinfo {volume} {294}},\ \bibinfo {pages} {1488} (\bibinfo {year} {2001})}\BibitemShut {NoStop}%
\bibitem [{\citenamefont {\ifmmode \check{Z}\else \v{Z}\fi{}uti\ifmmode~\acute{c}\else \'{c}\fi{}}\ \emph {et~al.}(2004)\citenamefont {\ifmmode \check{Z}\else \v{Z}\fi{}uti\ifmmode~\acute{c}\else \'{c}\fi{}}, \citenamefont {Fabian},\ and\ \citenamefont {Das~Sarma}}]{Zutic2004RevModPhys.76.323}%
  \BibitemOpen
  \bibfield  {author} {\bibinfo {author} {\bibfnamefont {I.}~\bibnamefont {\ifmmode \check{Z}\else \v{Z}\fi{}uti\ifmmode~\acute{c}\else \'{c}\fi{}}}, \bibinfo {author} {\bibfnamefont {J.}~\bibnamefont {Fabian}},\ and\ \bibinfo {author} {\bibfnamefont {S.}~\bibnamefont {Das~Sarma}},\ }\bibfield  {title} {\bibinfo {title} {Spintronics: Fundamentals and applications},\ }\href {https://doi.org/10.1103/RevModPhys.76.323} {\bibfield  {journal} {\bibinfo  {journal} {Rev. Mod. Phys.}\ }\textbf {\bibinfo {volume} {76}},\ \bibinfo {pages} {323} (\bibinfo {year} {2004})}\BibitemShut {NoStop}%
\bibitem [{\citenamefont {Puebla}\ \emph {et~al.}(2020)\citenamefont {Puebla}, \citenamefont {Kim}, \citenamefont {Kondou},\ and\ \citenamefont {Otani}}]{Puebla2020CommunMater.1.24}%
  \BibitemOpen
  \bibfield  {author} {\bibinfo {author} {\bibfnamefont {J.}~\bibnamefont {Puebla}}, \bibinfo {author} {\bibfnamefont {J.}~\bibnamefont {Kim}}, \bibinfo {author} {\bibfnamefont {K.}~\bibnamefont {Kondou}},\ and\ \bibinfo {author} {\bibfnamefont {Y.}~\bibnamefont {Otani}},\ }\bibfield  {title} {\bibinfo {title} {Spintronic devices for energy-efficient data storage and energy harvesting},\ }\href {https://www.nature.com/articles/s43246-020-0022-5} {\bibfield  {journal} {\bibinfo  {journal} {Commun. Mater.}\ }\textbf {\bibinfo {volume} {1}},\ \bibinfo {pages} {24} (\bibinfo {year} {2020})}\BibitemShut {NoStop}%
\bibitem [{\citenamefont {Parkin}\ \emph {et~al.}(2008)\citenamefont {Parkin}, \citenamefont {Hayashi},\ and\ \citenamefont {Thomas}}]{Parkin2008Science.320.190}%
  \BibitemOpen
  \bibfield  {author} {\bibinfo {author} {\bibfnamefont {S.~S.~P.}\ \bibnamefont {Parkin}}, \bibinfo {author} {\bibfnamefont {M.}~\bibnamefont {Hayashi}},\ and\ \bibinfo {author} {\bibfnamefont {L.}~\bibnamefont {Thomas}},\ }\bibfield  {title} {\bibinfo {title} {Magnetic domain-wall racetrack memory},\ }\href {https://doi.org/10.1126/science.1145799} {\bibfield  {journal} {\bibinfo  {journal} {Science}\ }\textbf {\bibinfo {volume} {320}},\ \bibinfo {pages} {190} (\bibinfo {year} {2008})}\BibitemShut {NoStop}%
\bibitem [{\citenamefont {Hirsch}(1999)}]{Hirsch1999PhysRevLett.83.1834}%
  \BibitemOpen
  \bibfield  {author} {\bibinfo {author} {\bibfnamefont {J.~E.}\ \bibnamefont {Hirsch}},\ }\bibfield  {title} {\bibinfo {title} {Spin hall effect},\ }\href {https://doi.org/10.1103/PhysRevLett.83.1834} {\bibfield  {journal} {\bibinfo  {journal} {Phys. Rev. Lett.}\ }\textbf {\bibinfo {volume} {83}},\ \bibinfo {pages} {1834} (\bibinfo {year} {1999})}\BibitemShut {NoStop}%
\bibitem [{\citenamefont {Kato}\ \emph {et~al.}(2004)\citenamefont {Kato}, \citenamefont {Myers}, \citenamefont {Gossard},\ and\ \citenamefont {Awschalom}}]{Kato2004Science.306.5703}%
  \BibitemOpen
  \bibfield  {author} {\bibinfo {author} {\bibfnamefont {Y.~K.}\ \bibnamefont {Kato}}, \bibinfo {author} {\bibfnamefont {R.~C.}\ \bibnamefont {Myers}}, \bibinfo {author} {\bibfnamefont {A.~C.}\ \bibnamefont {Gossard}},\ and\ \bibinfo {author} {\bibfnamefont {D.~D.}\ \bibnamefont {Awschalom}},\ }\bibfield  {title} {\bibinfo {title} {Observation of the spin hall effect in semiconductors},\ }\href {https://doi.org/10.1126/science.1105514} {\bibfield  {journal} {\bibinfo  {journal} {Science}\ }\textbf {\bibinfo {volume} {306}},\ \bibinfo {pages} {1910} (\bibinfo {year} {2004})}\BibitemShut {NoStop}%
\bibitem [{\citenamefont {Seki}\ \emph {et~al.}(2015)\citenamefont {Seki}, \citenamefont {Ideue}, \citenamefont {Kubota}, \citenamefont {Kozuka}, \citenamefont {Takagi}, \citenamefont {Nakamura}, \citenamefont {Kaneko}, \citenamefont {Kawasaki},\ and\ \citenamefont {Tokura}}]{Seki2015PhysRevLett115.266601}%
  \BibitemOpen
  \bibfield  {author} {\bibinfo {author} {\bibfnamefont {S.}~\bibnamefont {Seki}}, \bibinfo {author} {\bibfnamefont {T.}~\bibnamefont {Ideue}}, \bibinfo {author} {\bibfnamefont {M.}~\bibnamefont {Kubota}}, \bibinfo {author} {\bibfnamefont {Y.}~\bibnamefont {Kozuka}}, \bibinfo {author} {\bibfnamefont {R.}~\bibnamefont {Takagi}}, \bibinfo {author} {\bibfnamefont {M.}~\bibnamefont {Nakamura}}, \bibinfo {author} {\bibfnamefont {Y.}~\bibnamefont {Kaneko}}, \bibinfo {author} {\bibfnamefont {M.}~\bibnamefont {Kawasaki}},\ and\ \bibinfo {author} {\bibfnamefont {Y.}~\bibnamefont {Tokura}},\ }\bibfield  {title} {\bibinfo {title} {Thermal generation of spin current in an antiferromagnet},\ }\href {https://doi.org/10.1103/PhysRevLett.115.266601} {\bibfield  {journal} {\bibinfo  {journal} {Phys. Rev. Lett.}\ }\textbf {\bibinfo {volume} {115}},\ \bibinfo {pages} {266601} (\bibinfo {year} {2015})}\BibitemShut {NoStop}%
\bibitem [{\citenamefont {Rezende}\ \emph {et~al.}(2016)\citenamefont {Rezende}, \citenamefont {Rodr\'{\i}guez-Su\'arez},\ and\ \citenamefont {Azevedo}}]{Rezende2016PhysRevB93.014425}%
  \BibitemOpen
  \bibfield  {author} {\bibinfo {author} {\bibfnamefont {S.~M.}\ \bibnamefont {Rezende}}, \bibinfo {author} {\bibfnamefont {R.~L.}\ \bibnamefont {Rodr\'{\i}guez-Su\'arez}},\ and\ \bibinfo {author} {\bibfnamefont {A.}~\bibnamefont {Azevedo}},\ }\bibfield  {title} {\bibinfo {title} {Theory of the spin seebeck effect in antiferromagnets},\ }\href {https://doi.org/10.1103/PhysRevB.93.014425} {\bibfield  {journal} {\bibinfo  {journal} {Phys. Rev. B}\ }\textbf {\bibinfo {volume} {93}},\ \bibinfo {pages} {014425} (\bibinfo {year} {2016})}\BibitemShut {NoStop}%
\bibitem [{\citenamefont {Uchida}\ \emph {et~al.}(2010)\citenamefont {Uchida}, \citenamefont {Adachi}, \citenamefont {Ota}, \citenamefont {Nakayama}, \citenamefont {Maekawa},\ and\ \citenamefont {Saitoh}}]{Uchida2010ApplPhysLett.97.172505}%
  \BibitemOpen
  \bibfield  {author} {\bibinfo {author} {\bibfnamefont {K.-i.}\ \bibnamefont {Uchida}}, \bibinfo {author} {\bibfnamefont {H.}~\bibnamefont {Adachi}}, \bibinfo {author} {\bibfnamefont {T.}~\bibnamefont {Ota}}, \bibinfo {author} {\bibfnamefont {H.}~\bibnamefont {Nakayama}}, \bibinfo {author} {\bibfnamefont {S.}~\bibnamefont {Maekawa}},\ and\ \bibinfo {author} {\bibfnamefont {E.}~\bibnamefont {Saitoh}},\ }\bibfield  {title} {\bibinfo {title} {{Observation of longitudinal spin-Seebeck effect in magnetic insulators}},\ }\href {https://doi.org/10.1063/1.3507386} {\bibfield  {journal} {\bibinfo  {journal} {Appl. Phys. Lett.}\ }\textbf {\bibinfo {volume} {97}},\ \bibinfo {pages} {172505} (\bibinfo {year} {2010})}\BibitemShut {NoStop}%
\bibitem [{\citenamefont {Appelbaum}\ \emph {et~al.}(2007)\citenamefont {Appelbaum}, \citenamefont {Huang},\ and\ \citenamefont {Monsma}}]{Appelbaum2007Nature.447.7142}%
  \BibitemOpen
  \bibfield  {author} {\bibinfo {author} {\bibfnamefont {I.}~\bibnamefont {Appelbaum}}, \bibinfo {author} {\bibfnamefont {B.}~\bibnamefont {Huang}},\ and\ \bibinfo {author} {\bibfnamefont {D.~J.}\ \bibnamefont {Monsma}},\ }\bibfield  {title} {\bibinfo {title} {Electronic measurement and control of spin transport in silicon},\ }\href {https://www.nature.com/articles/nature05803} {\bibfield  {journal} {\bibinfo  {journal} {Nature (London)}\ }\textbf {\bibinfo {volume} {447}},\ \bibinfo {pages} {295} (\bibinfo {year} {2007})}\BibitemShut {NoStop}%
\bibitem [{\citenamefont {Lou}\ \emph {et~al.}(2007)\citenamefont {Lou}, \citenamefont {Adelmann}, \citenamefont {Crooker}, \citenamefont {Garlid}, \citenamefont {Zhang}, \citenamefont {Reddy}, \citenamefont {Flexner}, \citenamefont {Palmstr{\o}m},\ and\ \citenamefont {Crowell}}]{Lou2007NatPhys.3.197}%
  \BibitemOpen
  \bibfield  {author} {\bibinfo {author} {\bibfnamefont {X.}~\bibnamefont {Lou}}, \bibinfo {author} {\bibfnamefont {C.}~\bibnamefont {Adelmann}}, \bibinfo {author} {\bibfnamefont {S.~A.}\ \bibnamefont {Crooker}}, \bibinfo {author} {\bibfnamefont {E.~S.}\ \bibnamefont {Garlid}}, \bibinfo {author} {\bibfnamefont {J.}~\bibnamefont {Zhang}}, \bibinfo {author} {\bibfnamefont {K.~M.}\ \bibnamefont {Reddy}}, \bibinfo {author} {\bibfnamefont {S.~D.}\ \bibnamefont {Flexner}}, \bibinfo {author} {\bibfnamefont {C.~J.}\ \bibnamefont {Palmstr{\o}m}},\ and\ \bibinfo {author} {\bibfnamefont {P.~A.}\ \bibnamefont {Crowell}},\ }\bibfield  {title} {\bibinfo {title} {Electrical detection of spin transport in lateral ferromagnet--semiconductor devices},\ }\href {https://www.nature.com/articles/nphys543} {\bibfield  {journal} {\bibinfo  {journal} {Nat. Phys.}\ }\textbf {\bibinfo {volume} {3}},\ \bibinfo {pages} {197} (\bibinfo {year} {2007})}\BibitemShut {NoStop}%
\bibitem [{\citenamefont {Xiao}\ \emph {et~al.}(2010)\citenamefont {Xiao}, \citenamefont {Chang},\ and\ \citenamefont {Niu}}]{Xiao2010RevModPhys.82.1959}%
  \BibitemOpen
  \bibfield  {author} {\bibinfo {author} {\bibfnamefont {D.}~\bibnamefont {Xiao}}, \bibinfo {author} {\bibfnamefont {M.-C.}\ \bibnamefont {Chang}},\ and\ \bibinfo {author} {\bibfnamefont {Q.}~\bibnamefont {Niu}},\ }\bibfield  {title} {\bibinfo {title} {Berry phase effects on electronic properties},\ }\href {https://doi.org/10.1103/RevModPhys.82.1959} {\bibfield  {journal} {\bibinfo  {journal} {Rev. Mod. Phys.}\ }\textbf {\bibinfo {volume} {82}},\ \bibinfo {pages} {1959} (\bibinfo {year} {2010})}\BibitemShut {NoStop}%
\bibitem [{\citenamefont {Thonhauser}\ \emph {et~al.}(2005)\citenamefont {Thonhauser}, \citenamefont {Ceresoli}, \citenamefont {Vanderbilt},\ and\ \citenamefont {Resta}}]{Thonhauser2005PhysRevLett.95.137205}%
  \BibitemOpen
  \bibfield  {author} {\bibinfo {author} {\bibfnamefont {T.}~\bibnamefont {Thonhauser}}, \bibinfo {author} {\bibfnamefont {D.}~\bibnamefont {Ceresoli}}, \bibinfo {author} {\bibfnamefont {D.}~\bibnamefont {Vanderbilt}},\ and\ \bibinfo {author} {\bibfnamefont {R.}~\bibnamefont {Resta}},\ }\bibfield  {title} {\bibinfo {title} {Orbital magnetization in periodic insulators},\ }\href {https://doi.org/10.1103/PhysRevLett.95.137205} {\bibfield  {journal} {\bibinfo  {journal} {Phys. Rev. Lett.}\ }\textbf {\bibinfo {volume} {95}},\ \bibinfo {pages} {137205} (\bibinfo {year} {2005})}\BibitemShut {NoStop}%
\bibitem [{\citenamefont {Xiao}\ \emph {et~al.}(2005)\citenamefont {Xiao}, \citenamefont {Shi},\ and\ \citenamefont {Niu}}]{Xiao2005PhysRevLett.95.137204}%
  \BibitemOpen
  \bibfield  {author} {\bibinfo {author} {\bibfnamefont {D.}~\bibnamefont {Xiao}}, \bibinfo {author} {\bibfnamefont {J.}~\bibnamefont {Shi}},\ and\ \bibinfo {author} {\bibfnamefont {Q.}~\bibnamefont {Niu}},\ }\bibfield  {title} {\bibinfo {title} {Berry phase correction to electron density of states in solids},\ }\href {https://doi.org/10.1103/PhysRevLett.95.137204} {\bibfield  {journal} {\bibinfo  {journal} {Phys. Rev. Lett.}\ }\textbf {\bibinfo {volume} {95}},\ \bibinfo {pages} {137204} (\bibinfo {year} {2005})}\BibitemShut {NoStop}%
\bibitem [{\citenamefont {Go}\ \emph {et~al.}(2021)\citenamefont {Go}, \citenamefont {Jo}, \citenamefont {Lee}, \citenamefont {Kl{\"a}ui},\ and\ \citenamefont {Mokrousov}}]{Go2021EurophysLett.135.37001}%
  \BibitemOpen
  \bibfield  {author} {\bibinfo {author} {\bibfnamefont {D.}~\bibnamefont {Go}}, \bibinfo {author} {\bibfnamefont {D.}~\bibnamefont {Jo}}, \bibinfo {author} {\bibfnamefont {H.-W.}\ \bibnamefont {Lee}}, \bibinfo {author} {\bibfnamefont {M.}~\bibnamefont {Kl{\"a}ui}},\ and\ \bibinfo {author} {\bibfnamefont {Y.}~\bibnamefont {Mokrousov}},\ }\bibfield  {title} {\bibinfo {title} {Orbitronics: Orbital currents in solids},\ }\href {https://iopscience.iop.org/article/10.1209/0295-5075/ac2653} {\bibfield  {journal} {\bibinfo  {journal} {Europhys. Lett.}\ }\textbf {\bibinfo {volume} {135}},\ \bibinfo {pages} {37001} (\bibinfo {year} {2021})}\BibitemShut {NoStop}%
\bibitem [{\citenamefont {Jo}\ \emph {et~al.}(2018)\citenamefont {Jo}, \citenamefont {Go},\ and\ \citenamefont {Lee}}]{Jo2018PhysRevB.98.214405}%
  \BibitemOpen
  \bibfield  {author} {\bibinfo {author} {\bibfnamefont {D.}~\bibnamefont {Jo}}, \bibinfo {author} {\bibfnamefont {D.}~\bibnamefont {Go}},\ and\ \bibinfo {author} {\bibfnamefont {H.-W.}\ \bibnamefont {Lee}},\ }\bibfield  {title} {\bibinfo {title} {Gigantic intrinsic orbital hall effects in weakly spin-orbit coupled metals},\ }\href {https://doi.org/10.1103/PhysRevB.98.214405} {\bibfield  {journal} {\bibinfo  {journal} {Phys. Rev. B}\ }\textbf {\bibinfo {volume} {98}},\ \bibinfo {pages} {214405} (\bibinfo {year} {2018})}\BibitemShut {NoStop}%
\bibitem [{\citenamefont {Kontani}\ \emph {et~al.}(2009)\citenamefont {Kontani}, \citenamefont {Tanaka}, \citenamefont {Hirashima}, \citenamefont {Yamada},\ and\ \citenamefont {Inoue}}]{Kontani2009PhysRevLett.102.016601}%
  \BibitemOpen
  \bibfield  {author} {\bibinfo {author} {\bibfnamefont {H.}~\bibnamefont {Kontani}}, \bibinfo {author} {\bibfnamefont {T.}~\bibnamefont {Tanaka}}, \bibinfo {author} {\bibfnamefont {D.~S.}\ \bibnamefont {Hirashima}}, \bibinfo {author} {\bibfnamefont {K.}~\bibnamefont {Yamada}},\ and\ \bibinfo {author} {\bibfnamefont {J.}~\bibnamefont {Inoue}},\ }\bibfield  {title} {\bibinfo {title} {Giant orbital hall effect in transition metals: Origin of large spin and anomalous hall effects},\ }\href {https://doi.org/10.1103/PhysRevLett.102.016601} {\bibfield  {journal} {\bibinfo  {journal} {Phys. Rev. Lett.}\ }\textbf {\bibinfo {volume} {102}},\ \bibinfo {pages} {016601} (\bibinfo {year} {2009})}\BibitemShut {NoStop}%
\bibitem [{\citenamefont {Go}\ \emph {et~al.}(2018)\citenamefont {Go}, \citenamefont {Jo}, \citenamefont {Kim},\ and\ \citenamefont {Lee}}]{Go2018PhysRevLett.121.086602}%
  \BibitemOpen
  \bibfield  {author} {\bibinfo {author} {\bibfnamefont {D.}~\bibnamefont {Go}}, \bibinfo {author} {\bibfnamefont {D.}~\bibnamefont {Jo}}, \bibinfo {author} {\bibfnamefont {C.}~\bibnamefont {Kim}},\ and\ \bibinfo {author} {\bibfnamefont {H.-W.}\ \bibnamefont {Lee}},\ }\bibfield  {title} {\bibinfo {title} {Intrinsic spin and orbital hall effects from orbital texture},\ }\href {https://doi.org/10.1103/PhysRevLett.121.086602} {\bibfield  {journal} {\bibinfo  {journal} {Phys. Rev. Lett.}\ }\textbf {\bibinfo {volume} {121}},\ \bibinfo {pages} {086602} (\bibinfo {year} {2018})}\BibitemShut {NoStop}%
\bibitem [{\citenamefont {Kruglyak}\ \emph {et~al.}(2010)\citenamefont {Kruglyak}, \citenamefont {Demokritov},\ and\ \citenamefont {Grundler}}]{Kruglyak2010JPhysD43.264001}%
  \BibitemOpen
  \bibfield  {author} {\bibinfo {author} {\bibfnamefont {V.}~\bibnamefont {Kruglyak}}, \bibinfo {author} {\bibfnamefont {S.}~\bibnamefont {Demokritov}},\ and\ \bibinfo {author} {\bibfnamefont {D.}~\bibnamefont {Grundler}},\ }\bibfield  {title} {\bibinfo {title} {Magnonics},\ }\href {https://iopscience.iop.org/article/10.1088/0022-3727/43/26/260301} {\bibfield  {journal} {\bibinfo  {journal} {J. Phys. D}\ }\textbf {\bibinfo {volume} {43}},\ \bibinfo {pages} {264001} (\bibinfo {year} {2010})}\BibitemShut {NoStop}%
\bibitem [{\citenamefont {Chumak}\ \emph {et~al.}(2015)\citenamefont {Chumak}, \citenamefont {Vasyuchka}, \citenamefont {Serga},\ and\ \citenamefont {Hillebrands}}]{Chumak2015NatPhys.11.453}%
  \BibitemOpen
  \bibfield  {author} {\bibinfo {author} {\bibfnamefont {A.~V.}\ \bibnamefont {Chumak}}, \bibinfo {author} {\bibfnamefont {V.~I.}\ \bibnamefont {Vasyuchka}}, \bibinfo {author} {\bibfnamefont {A.~A.}\ \bibnamefont {Serga}},\ and\ \bibinfo {author} {\bibfnamefont {B.}~\bibnamefont {Hillebrands}},\ }\bibfield  {title} {\bibinfo {title} {Magnon spintronics},\ }\href {https://www.nature.com/articles/nphys3347} {\bibfield  {journal} {\bibinfo  {journal} {Nat. Phys.}\ }\textbf {\bibinfo {volume} {11}},\ \bibinfo {pages} {453} (\bibinfo {year} {2015})}\BibitemShut {NoStop}%
\bibitem [{\citenamefont {Bozhko}\ \emph {et~al.}(2020)\citenamefont {Bozhko}, \citenamefont {Vasyuchka}, \citenamefont {Chumak},\ and\ \citenamefont {Serga}}]{Bozhko2020LowTempPhys.46.383}%
  \BibitemOpen
  \bibfield  {author} {\bibinfo {author} {\bibfnamefont {D.~A.}\ \bibnamefont {Bozhko}}, \bibinfo {author} {\bibfnamefont {V.~I.}\ \bibnamefont {Vasyuchka}}, \bibinfo {author} {\bibfnamefont {A.~V.}\ \bibnamefont {Chumak}},\ and\ \bibinfo {author} {\bibfnamefont {A.~A.}\ \bibnamefont {Serga}},\ }\bibfield  {title} {\bibinfo {title} {{Magnon-phonon interactions in magnon spintronics (Review article)}},\ }\href {https://doi.org/10.1063/10.0000872} {\bibfield  {journal} {\bibinfo  {journal} {Low Temp. Phys.}\ }\textbf {\bibinfo {volume} {46}},\ \bibinfo {pages} {383} (\bibinfo {year} {2020})}\BibitemShut {NoStop}%
\bibitem [{\citenamefont {Serga}\ \emph {et~al.}(2010)\citenamefont {Serga}, \citenamefont {Chumak},\ and\ \citenamefont {Hillebrands}}]{Serga2010JPhysD.43.264002}%
  \BibitemOpen
  \bibfield  {author} {\bibinfo {author} {\bibfnamefont {A.}~\bibnamefont {Serga}}, \bibinfo {author} {\bibfnamefont {A.}~\bibnamefont {Chumak}},\ and\ \bibinfo {author} {\bibfnamefont {B.}~\bibnamefont {Hillebrands}},\ }\bibfield  {title} {\bibinfo {title} {Yig magnonics},\ }\href {https://iopscience.iop.org/article/10.1088/0022-3727/43/26/264002} {\bibfield  {journal} {\bibinfo  {journal} {J. Phys. D}\ }\textbf {\bibinfo {volume} {43}},\ \bibinfo {pages} {264002} (\bibinfo {year} {2010})}\BibitemShut {NoStop}%
\bibitem [{\citenamefont {Chumak}\ \emph {et~al.}(2014)\citenamefont {Chumak}, \citenamefont {Serga},\ and\ \citenamefont {Hillebrands}}]{Chumak2014NatCommun.5.4700}%
  \BibitemOpen
  \bibfield  {author} {\bibinfo {author} {\bibfnamefont {A.~V.}\ \bibnamefont {Chumak}}, \bibinfo {author} {\bibfnamefont {A.~A.}\ \bibnamefont {Serga}},\ and\ \bibinfo {author} {\bibfnamefont {B.}~\bibnamefont {Hillebrands}},\ }\bibfield  {title} {\bibinfo {title} {Magnon transistor for all-magnon data processing},\ }\href {https://doi.org/10.1038/ncomms5700} {\bibfield  {journal} {\bibinfo  {journal} {Nat. Commun.}\ }\textbf {\bibinfo {volume} {5}},\ \bibinfo {pages} {4700} (\bibinfo {year} {2014})}\BibitemShut {NoStop}%
\bibitem [{\citenamefont {Cornelissen}\ \emph {et~al.}(2015)\citenamefont {Cornelissen}, \citenamefont {Liu}, \citenamefont {Duine}, \citenamefont {Youssef},\ and\ \citenamefont {Van~Wees}}]{Cornelissen2015NatPhys.11.1022}%
  \BibitemOpen
  \bibfield  {author} {\bibinfo {author} {\bibfnamefont {L.}~\bibnamefont {Cornelissen}}, \bibinfo {author} {\bibfnamefont {J.}~\bibnamefont {Liu}}, \bibinfo {author} {\bibfnamefont {R.}~\bibnamefont {Duine}}, \bibinfo {author} {\bibfnamefont {J.~B.}\ \bibnamefont {Youssef}},\ and\ \bibinfo {author} {\bibfnamefont {B.}~\bibnamefont {Van~Wees}},\ }\bibfield  {title} {\bibinfo {title} {Long-distance transport of magnon spin information in a magnetic insulator at room temperature},\ }\href {https://www.nature.com/articles/nphys3465} {\bibfield  {journal} {\bibinfo  {journal} {Nat. Phys.}\ }\textbf {\bibinfo {volume} {11}},\ \bibinfo {pages} {1022} (\bibinfo {year} {2015})}\BibitemShut {NoStop}%
\bibitem [{\citenamefont {Kampfrath}\ \emph {et~al.}(2011)\citenamefont {Kampfrath}, \citenamefont {Sell}, \citenamefont {Klatt}, \citenamefont {Pashkin}, \citenamefont {M{\"a}hrlein}, \citenamefont {Dekorsy}, \citenamefont {Wolf}, \citenamefont {Fiebig}, \citenamefont {Leitenstorfer},\ and\ \citenamefont {Huber}}]{Kampfrath2011NatPhoton.5.31}%
  \BibitemOpen
  \bibfield  {author} {\bibinfo {author} {\bibfnamefont {T.}~\bibnamefont {Kampfrath}}, \bibinfo {author} {\bibfnamefont {A.}~\bibnamefont {Sell}}, \bibinfo {author} {\bibfnamefont {G.}~\bibnamefont {Klatt}}, \bibinfo {author} {\bibfnamefont {A.}~\bibnamefont {Pashkin}}, \bibinfo {author} {\bibfnamefont {S.}~\bibnamefont {M{\"a}hrlein}}, \bibinfo {author} {\bibfnamefont {T.}~\bibnamefont {Dekorsy}}, \bibinfo {author} {\bibfnamefont {M.}~\bibnamefont {Wolf}}, \bibinfo {author} {\bibfnamefont {M.}~\bibnamefont {Fiebig}}, \bibinfo {author} {\bibfnamefont {A.}~\bibnamefont {Leitenstorfer}},\ and\ \bibinfo {author} {\bibfnamefont {R.}~\bibnamefont {Huber}},\ }\bibfield  {title} {\bibinfo {title} {Coherent terahertz control of antiferromagnetic spin waves},\ }\href {https://www.nature.com/articles/nphoton.2010.259} {\bibfield  {journal} {\bibinfo  {journal} {Nat. Photon.}\ }\textbf {\bibinfo {volume} {5}},\ \bibinfo {pages} {31} (\bibinfo {year} {2011})}\BibitemShut {NoStop}%
\bibitem [{\citenamefont {Li}\ \emph {et~al.}(2020{\natexlab{a}})\citenamefont {Li}, \citenamefont {Wilson}, \citenamefont {Cheng}, \citenamefont {Lohmann}, \citenamefont {Kavand}, \citenamefont {Yuan}, \citenamefont {Aldosary}, \citenamefont {Agladze}, \citenamefont {Wei}, \citenamefont {Sherwin} \emph {et~al.}}]{Li2020Nature.578.70}%
  \BibitemOpen
  \bibfield  {author} {\bibinfo {author} {\bibfnamefont {J.}~\bibnamefont {Li}}, \bibinfo {author} {\bibfnamefont {C.~B.}\ \bibnamefont {Wilson}}, \bibinfo {author} {\bibfnamefont {R.}~\bibnamefont {Cheng}}, \bibinfo {author} {\bibfnamefont {M.}~\bibnamefont {Lohmann}}, \bibinfo {author} {\bibfnamefont {M.}~\bibnamefont {Kavand}}, \bibinfo {author} {\bibfnamefont {W.}~\bibnamefont {Yuan}}, \bibinfo {author} {\bibfnamefont {M.}~\bibnamefont {Aldosary}}, \bibinfo {author} {\bibfnamefont {N.}~\bibnamefont {Agladze}}, \bibinfo {author} {\bibfnamefont {P.}~\bibnamefont {Wei}}, \bibinfo {author} {\bibfnamefont {M.~S.}\ \bibnamefont {Sherwin}}, \emph {et~al.},\ }\bibfield  {title} {\bibinfo {title} {Spin current from sub-terahertz-generated antiferromagnetic magnons},\ }\href {https://www.nature.com/articles/s41586-020-1950-4} {\bibfield  {journal} {\bibinfo  {journal} {Nature (London)}\ }\textbf {\bibinfo {volume} {578}},\ \bibinfo {pages} {70} (\bibinfo {year} {2020}{\natexlab{a}})}\BibitemShut {NoStop}%
\bibitem [{\citenamefont {Neumann}\ \emph {et~al.}(2020)\citenamefont {Neumann}, \citenamefont {Mook}, \citenamefont {Henk},\ and\ \citenamefont {Mertig}}]{Neumann2020PhysRevLett.125.117209}%
  \BibitemOpen
  \bibfield  {author} {\bibinfo {author} {\bibfnamefont {R.~R.}\ \bibnamefont {Neumann}}, \bibinfo {author} {\bibfnamefont {A.}~\bibnamefont {Mook}}, \bibinfo {author} {\bibfnamefont {J.}~\bibnamefont {Henk}},\ and\ \bibinfo {author} {\bibfnamefont {I.}~\bibnamefont {Mertig}},\ }\bibfield  {title} {\bibinfo {title} {Orbital magnetic moment of magnons},\ }\href {https://doi.org/10.1103/PhysRevLett.125.117209} {\bibfield  {journal} {\bibinfo  {journal} {Phys. Rev. Lett.}\ }\textbf {\bibinfo {volume} {125}},\ \bibinfo {pages} {117209} (\bibinfo {year} {2020})}\BibitemShut {NoStop}%
\bibitem [{\citenamefont {Fishman}\ \emph {et~al.}(2022{\natexlab{a}})\citenamefont {Fishman}, \citenamefont {Gardner},\ and\ \citenamefont {Okamoto}}]{Fishman2022PhysRevLett.129.167202}%
  \BibitemOpen
  \bibfield  {author} {\bibinfo {author} {\bibfnamefont {R.~S.}\ \bibnamefont {Fishman}}, \bibinfo {author} {\bibfnamefont {J.~S.}\ \bibnamefont {Gardner}},\ and\ \bibinfo {author} {\bibfnamefont {S.}~\bibnamefont {Okamoto}},\ }\bibfield  {title} {\bibinfo {title} {Orbital angular momentum of magnons in collinear magnets},\ }\href {https://doi.org/10.1103/PhysRevLett.129.167202} {\bibfield  {journal} {\bibinfo  {journal} {Phys. Rev. Lett.}\ }\textbf {\bibinfo {volume} {129}},\ \bibinfo {pages} {167202} (\bibinfo {year} {2022}{\natexlab{a}})}\BibitemShut {NoStop}%
\bibitem [{\citenamefont {Fishman}\ \emph {et~al.}(2022{\natexlab{b}})\citenamefont {Fishman}, \citenamefont {Lindsay},\ and\ \citenamefont {Okamoto}}]{Fishman2022JCdsMaterPhys.51.015801}%
  \BibitemOpen
  \bibfield  {author} {\bibinfo {author} {\bibfnamefont {R.~S.}\ \bibnamefont {Fishman}}, \bibinfo {author} {\bibfnamefont {L.}~\bibnamefont {Lindsay}},\ and\ \bibinfo {author} {\bibfnamefont {S.}~\bibnamefont {Okamoto}},\ }\bibfield  {title} {\bibinfo {title} {Exact results for the orbital angular momentum of magnons on honeycomb lattices},\ }\href {https://iopscience.iop.org/article/10.1088/1361-648X/ac9a28} {\bibfield  {journal} {\bibinfo  {journal} {J. Condens. Matter Phys.}\ }\textbf {\bibinfo {volume} {51}},\ \bibinfo {pages} {015801} (\bibinfo {year} {2022}{\natexlab{b}})}\BibitemShut {NoStop}%
\bibitem [{\citenamefont {Fishman}(2023)}]{Fishman2023PhysRevB.107.214434}%
  \BibitemOpen
  \bibfield  {author} {\bibinfo {author} {\bibfnamefont {R.~S.}\ \bibnamefont {Fishman}},\ }\bibfield  {title} {\bibinfo {title} {Gauge-invariant measure of the magnon orbital angular momentum},\ }\href {https://doi.org/10.1103/PhysRevB.107.214434} {\bibfield  {journal} {\bibinfo  {journal} {Phys. Rev. B}\ }\textbf {\bibinfo {volume} {107}},\ \bibinfo {pages} {214434} (\bibinfo {year} {2023})}\BibitemShut {NoStop}%
\bibitem [{\citenamefont {Go}\ \emph {et~al.}(2024)\citenamefont {Go}, \citenamefont {An}, \citenamefont {Lee},\ and\ \citenamefont {Kim}}]{Go2024NanoLett.24.5968}%
  \BibitemOpen
  \bibfield  {author} {\bibinfo {author} {\bibfnamefont {G.}~\bibnamefont {Go}}, \bibinfo {author} {\bibfnamefont {D.}~\bibnamefont {An}}, \bibinfo {author} {\bibfnamefont {H.-W.}\ \bibnamefont {Lee}},\ and\ \bibinfo {author} {\bibfnamefont {S.~K.}\ \bibnamefont {Kim}},\ }\bibfield  {title} {\bibinfo {title} {Magnon orbital nernst effect in honeycomb antiferromagnets without spin–orbit coupling},\ }\href {https://doi.org/10.1021/acs.nanolett.4c00430} {\bibfield  {journal} {\bibinfo  {journal} {Nano Letters}\ }\textbf {\bibinfo {volume} {24}},\ \bibinfo {pages} {5968} (\bibinfo {year} {2024})}\BibitemShut {NoStop}%
\bibitem [{\citenamefont {To}\ \emph {et~al.}(2024)\citenamefont {To}, \citenamefont {Garcia-Gaitan}, \citenamefont {Ren}, \citenamefont {Zide}, \citenamefont {Xiao}, \citenamefont {Nikolić}, \citenamefont {Bryant},\ and\ \citenamefont {Doty}}]{To2024ArXiv.16004}%
  \BibitemOpen
  \bibfield  {author} {\bibinfo {author} {\bibfnamefont {D.~Q.}\ \bibnamefont {To}}, \bibinfo {author} {\bibfnamefont {F.}~\bibnamefont {Garcia-Gaitan}}, \bibinfo {author} {\bibfnamefont {Y.}~\bibnamefont {Ren}}, \bibinfo {author} {\bibfnamefont {J.~M.~O.}\ \bibnamefont {Zide}}, \bibinfo {author} {\bibfnamefont {J.~Q.}\ \bibnamefont {Xiao}}, \bibinfo {author} {\bibfnamefont {B.~K.}\ \bibnamefont {Nikolić}}, \bibinfo {author} {\bibfnamefont {G.~W.}\ \bibnamefont {Bryant}},\ and\ \bibinfo {author} {\bibfnamefont {M.~F.}\ \bibnamefont {Doty}},\ }\href@noop {} {\bibinfo {title} {Theory of electric polarization induced by magnon transport in two-dimensional honeycomb antiferromagnets}} (\bibinfo {year} {2024}),\ \Eprint {https://arxiv.org/abs/2407.16004} {arXiv:2407.16004} \BibitemShut {NoStop}%
\bibitem [{\citenamefont {Kitaev}(2006)}]{Kitaev2006AnnPhys.321.2}%
  \BibitemOpen
  \bibfield  {author} {\bibinfo {author} {\bibfnamefont {A.}~\bibnamefont {Kitaev}},\ }\bibfield  {title} {\bibinfo {title} {Anyons in an exactly solved model and beyond},\ }\href {https://doi.org/https://doi.org/10.1016/j.aop.2005.10.005} {\bibfield  {journal} {\bibinfo  {journal} {Ann. Phys.}\ }\textbf {\bibinfo {volume} {321}},\ \bibinfo {pages} {2} (\bibinfo {year} {2006})},\ \bibinfo {note} {january Special Issue}\BibitemShut {NoStop}%
\bibitem [{\citenamefont {Jackeli}\ and\ \citenamefont {Khaliullin}(2009)}]{Jackeli2009PhysRevLett.102.017205}%
  \BibitemOpen
  \bibfield  {author} {\bibinfo {author} {\bibfnamefont {G.}~\bibnamefont {Jackeli}}\ and\ \bibinfo {author} {\bibfnamefont {G.}~\bibnamefont {Khaliullin}},\ }\bibfield  {title} {\bibinfo {title} {Mott insulators in the strong spin-orbit coupling limit: From heisenberg to a quantum compass and kitaev models},\ }\href {https://doi.org/10.1103/PhysRevLett.102.017205} {\bibfield  {journal} {\bibinfo  {journal} {Phys. Rev. Lett.}\ }\textbf {\bibinfo {volume} {102}},\ \bibinfo {pages} {017205} (\bibinfo {year} {2009})}\BibitemShut {NoStop}%
\bibitem [{\citenamefont {Kim}\ \emph {et~al.}(2016)\citenamefont {Kim}, \citenamefont {Ochoa}, \citenamefont {Zarzuela},\ and\ \citenamefont {Tserkovnyak}}]{Kim2016PhysRevLett.117.227201}%
  \BibitemOpen
  \bibfield  {author} {\bibinfo {author} {\bibfnamefont {S.~K.}\ \bibnamefont {Kim}}, \bibinfo {author} {\bibfnamefont {H.}~\bibnamefont {Ochoa}}, \bibinfo {author} {\bibfnamefont {R.}~\bibnamefont {Zarzuela}},\ and\ \bibinfo {author} {\bibfnamefont {Y.}~\bibnamefont {Tserkovnyak}},\ }\bibfield  {title} {\bibinfo {title} {Realization of the haldane-kane-mele model in a system of localized spins},\ }\href {https://doi.org/10.1103/PhysRevLett.117.227201} {\bibfield  {journal} {\bibinfo  {journal} {Phys. Rev. Lett.}\ }\textbf {\bibinfo {volume} {117}},\ \bibinfo {pages} {227201} (\bibinfo {year} {2016})}\BibitemShut {NoStop}%
\bibitem [{\citenamefont {Chen}\ \emph {et~al.}(2018)\citenamefont {Chen}, \citenamefont {Chung}, \citenamefont {Gao}, \citenamefont {Chen}, \citenamefont {Stone}, \citenamefont {Kolesnikov}, \citenamefont {Huang},\ and\ \citenamefont {Dai}}]{Chen2018PhysRevX.8.041028}%
  \BibitemOpen
  \bibfield  {author} {\bibinfo {author} {\bibfnamefont {L.}~\bibnamefont {Chen}}, \bibinfo {author} {\bibfnamefont {J.-H.}\ \bibnamefont {Chung}}, \bibinfo {author} {\bibfnamefont {B.}~\bibnamefont {Gao}}, \bibinfo {author} {\bibfnamefont {T.}~\bibnamefont {Chen}}, \bibinfo {author} {\bibfnamefont {M.~B.}\ \bibnamefont {Stone}}, \bibinfo {author} {\bibfnamefont {A.~I.}\ \bibnamefont {Kolesnikov}}, \bibinfo {author} {\bibfnamefont {Q.}~\bibnamefont {Huang}},\ and\ \bibinfo {author} {\bibfnamefont {P.}~\bibnamefont {Dai}},\ }\bibfield  {title} {\bibinfo {title} {Topological spin excitations in honeycomb ferromagnet ${\mathrm{cri}}_{3}$},\ }\href {https://doi.org/10.1103/PhysRevX.8.041028} {\bibfield  {journal} {\bibinfo  {journal} {Phys. Rev. X}\ }\textbf {\bibinfo {volume} {8}},\ \bibinfo {pages} {041028} (\bibinfo {year} {2018})}\BibitemShut {NoStop}%
\bibitem [{\citenamefont {Chen}\ \emph {et~al.}(2020)\citenamefont {Chen}, \citenamefont {Chung}, \citenamefont {Chen}, \citenamefont {Duan}, \citenamefont {Schneidewind}, \citenamefont {Radelytskyi}, \citenamefont {Voneshen}, \citenamefont {Ewings}, \citenamefont {Stone}, \citenamefont {Kolesnikov}, \citenamefont {Winn}, \citenamefont {Chi}, \citenamefont {Mole}, \citenamefont {Yu}, \citenamefont {Gao},\ and\ \citenamefont {Dai}}]{Chen2020PhysRevB.101.134418}%
  \BibitemOpen
  \bibfield  {author} {\bibinfo {author} {\bibfnamefont {L.}~\bibnamefont {Chen}}, \bibinfo {author} {\bibfnamefont {J.-H.}\ \bibnamefont {Chung}}, \bibinfo {author} {\bibfnamefont {T.}~\bibnamefont {Chen}}, \bibinfo {author} {\bibfnamefont {C.}~\bibnamefont {Duan}}, \bibinfo {author} {\bibfnamefont {A.}~\bibnamefont {Schneidewind}}, \bibinfo {author} {\bibfnamefont {I.}~\bibnamefont {Radelytskyi}}, \bibinfo {author} {\bibfnamefont {D.~J.}\ \bibnamefont {Voneshen}}, \bibinfo {author} {\bibfnamefont {R.~A.}\ \bibnamefont {Ewings}}, \bibinfo {author} {\bibfnamefont {M.~B.}\ \bibnamefont {Stone}}, \bibinfo {author} {\bibfnamefont {A.~I.}\ \bibnamefont {Kolesnikov}}, \bibinfo {author} {\bibfnamefont {B.}~\bibnamefont {Winn}}, \bibinfo {author} {\bibfnamefont {S.}~\bibnamefont {Chi}}, \bibinfo {author} {\bibfnamefont {R.~A.}\ \bibnamefont {Mole}}, \bibinfo {author} {\bibfnamefont {D.~H.}\ \bibnamefont {Yu}}, \bibinfo {author} {\bibfnamefont {B.}~\bibnamefont {Gao}},\ and\ \bibinfo {author} {\bibfnamefont
  {P.}~\bibnamefont {Dai}},\ }\bibfield  {title} {\bibinfo {title} {Magnetic anisotropy in ferromagnetic ${\mathrm{cri}}_{3}$},\ }\href {https://doi.org/10.1103/PhysRevB.101.134418} {\bibfield  {journal} {\bibinfo  {journal} {Phys. Rev. B}\ }\textbf {\bibinfo {volume} {101}},\ \bibinfo {pages} {134418} (\bibinfo {year} {2020})}\BibitemShut {NoStop}%
\bibitem [{\citenamefont {Lee}\ \emph {et~al.}(2020)\citenamefont {Lee}, \citenamefont {Utermohlen}, \citenamefont {Weber}, \citenamefont {Hwang}, \citenamefont {Zhang}, \citenamefont {van Tol}, \citenamefont {Goldberger}, \citenamefont {Trivedi},\ and\ \citenamefont {Hammel}}]{Lee2020PhysRevLett.124.017201}%
  \BibitemOpen
  \bibfield  {author} {\bibinfo {author} {\bibfnamefont {I.}~\bibnamefont {Lee}}, \bibinfo {author} {\bibfnamefont {F.~G.}\ \bibnamefont {Utermohlen}}, \bibinfo {author} {\bibfnamefont {D.}~\bibnamefont {Weber}}, \bibinfo {author} {\bibfnamefont {K.}~\bibnamefont {Hwang}}, \bibinfo {author} {\bibfnamefont {C.}~\bibnamefont {Zhang}}, \bibinfo {author} {\bibfnamefont {J.}~\bibnamefont {van Tol}}, \bibinfo {author} {\bibfnamefont {J.~E.}\ \bibnamefont {Goldberger}}, \bibinfo {author} {\bibfnamefont {N.}~\bibnamefont {Trivedi}},\ and\ \bibinfo {author} {\bibfnamefont {P.~C.}\ \bibnamefont {Hammel}},\ }\bibfield  {title} {\bibinfo {title} {Fundamental spin interactions underlying the magnetic anisotropy in the kitaev ferromagnet ${\mathrm{cri}}_{3}$},\ }\href {https://doi.org/10.1103/PhysRevLett.124.017201} {\bibfield  {journal} {\bibinfo  {journal} {Phys. Rev. Lett.}\ }\textbf {\bibinfo {volume} {124}},\ \bibinfo {pages} {017201} (\bibinfo {year} {2020})}\BibitemShut {NoStop}%
\bibitem [{\citenamefont {Chen}\ \emph {et~al.}(2021)\citenamefont {Chen}, \citenamefont {Chung}, \citenamefont {Stone}, \citenamefont {Kolesnikov}, \citenamefont {Winn}, \citenamefont {Garlea}, \citenamefont {Abernathy}, \citenamefont {Gao}, \citenamefont {Augustin}, \citenamefont {Santos},\ and\ \citenamefont {Dai}}]{Chen2021PhysRevX.11.031047}%
  \BibitemOpen
  \bibfield  {author} {\bibinfo {author} {\bibfnamefont {L.}~\bibnamefont {Chen}}, \bibinfo {author} {\bibfnamefont {J.-H.}\ \bibnamefont {Chung}}, \bibinfo {author} {\bibfnamefont {M.~B.}\ \bibnamefont {Stone}}, \bibinfo {author} {\bibfnamefont {A.~I.}\ \bibnamefont {Kolesnikov}}, \bibinfo {author} {\bibfnamefont {B.}~\bibnamefont {Winn}}, \bibinfo {author} {\bibfnamefont {V.~O.}\ \bibnamefont {Garlea}}, \bibinfo {author} {\bibfnamefont {D.~L.}\ \bibnamefont {Abernathy}}, \bibinfo {author} {\bibfnamefont {B.}~\bibnamefont {Gao}}, \bibinfo {author} {\bibfnamefont {M.}~\bibnamefont {Augustin}}, \bibinfo {author} {\bibfnamefont {E.~J.~G.}\ \bibnamefont {Santos}},\ and\ \bibinfo {author} {\bibfnamefont {P.}~\bibnamefont {Dai}},\ }\bibfield  {title} {\bibinfo {title} {Magnetic field effect on topological spin excitations in ${\mathrm{cri}}_{3}$},\ }\href {https://doi.org/10.1103/PhysRevX.11.031047} {\bibfield  {journal} {\bibinfo  {journal} {Phys. Rev. X}\ }\textbf {\bibinfo {volume} {11}},\ \bibinfo {pages}
  {031047} (\bibinfo {year} {2021})}\BibitemShut {NoStop}%
\bibitem [{\citenamefont {Zhang}\ \emph {et~al.}(2021{\natexlab{a}})\citenamefont {Zhang}, \citenamefont {Zhu}, \citenamefont {Go}, \citenamefont {Lux}, \citenamefont {dos Santos}, \citenamefont {Lounis}, \citenamefont {Su}, \citenamefont {Bl\"ugel},\ and\ \citenamefont {Mokrousov}}]{Zhang2021PhysRevB.103.134414}%
  \BibitemOpen
  \bibfield  {author} {\bibinfo {author} {\bibfnamefont {L.-C.}\ \bibnamefont {Zhang}}, \bibinfo {author} {\bibfnamefont {F.}~\bibnamefont {Zhu}}, \bibinfo {author} {\bibfnamefont {D.}~\bibnamefont {Go}}, \bibinfo {author} {\bibfnamefont {F.~R.}\ \bibnamefont {Lux}}, \bibinfo {author} {\bibfnamefont {F.~J.}\ \bibnamefont {dos Santos}}, \bibinfo {author} {\bibfnamefont {S.}~\bibnamefont {Lounis}}, \bibinfo {author} {\bibfnamefont {Y.}~\bibnamefont {Su}}, \bibinfo {author} {\bibfnamefont {S.}~\bibnamefont {Bl\"ugel}},\ and\ \bibinfo {author} {\bibfnamefont {Y.}~\bibnamefont {Mokrousov}},\ }\bibfield  {title} {\bibinfo {title} {Interplay of dzyaloshinskii-moriya and kitaev interactions for magnonic properties of heisenberg-kitaev honeycomb ferromagnets},\ }\href {https://doi.org/10.1103/PhysRevB.103.134414} {\bibfield  {journal} {\bibinfo  {journal} {Phys. Rev. B}\ }\textbf {\bibinfo {volume} {103}},\ \bibinfo {pages} {134414} (\bibinfo {year} {2021}{\natexlab{a}})}\BibitemShut {NoStop}%
\bibitem [{\citenamefont {Brehm}\ \emph {et~al.}(2024)\citenamefont {Brehm}, \citenamefont {Sobieszczyk},\ and\ \citenamefont {Qaiumzadeh}}]{Brehm2024ArXiv.15964}%
  \BibitemOpen
  \bibfield  {author} {\bibinfo {author} {\bibfnamefont {V.}~\bibnamefont {Brehm}}, \bibinfo {author} {\bibfnamefont {P.}~\bibnamefont {Sobieszczyk}},\ and\ \bibinfo {author} {\bibfnamefont {A.}~\bibnamefont {Qaiumzadeh}},\ }\href@noop {} {\bibinfo {title} {Intrinsic spin nernst effect and chiral edge modes in van der waals ferromagnetic insulators: Dzyaloshinskii-moriya vs. kitaev interactions}} (\bibinfo {year} {2024}),\ \Eprint {https://arxiv.org/abs/2409.15964} {arXiv:2409.15964} \BibitemShut {NoStop}%
\bibitem [{\citenamefont {Holstein}\ and\ \citenamefont {Primakoff}(1940)}]{Holstein1940PhysRev.58.1098}%
  \BibitemOpen
  \bibfield  {author} {\bibinfo {author} {\bibfnamefont {T.}~\bibnamefont {Holstein}}\ and\ \bibinfo {author} {\bibfnamefont {H.}~\bibnamefont {Primakoff}},\ }\bibfield  {title} {\bibinfo {title} {Field dependence of the intrinsic domain magnetization of a ferromagnet},\ }\href {https://doi.org/10.1103/PhysRev.58.1098} {\bibfield  {journal} {\bibinfo  {journal} {Phys. Rev.}\ }\textbf {\bibinfo {volume} {58}},\ \bibinfo {pages} {1098} (\bibinfo {year} {1940})}\BibitemShut {NoStop}%
\bibitem [{\citenamefont {Zhang}\ \emph {et~al.}(2021{\natexlab{b}})\citenamefont {Zhang}, \citenamefont {Chern},\ and\ \citenamefont {Kim}}]{Zhang2021PhysRevB.103.174402}%
  \BibitemOpen
  \bibfield  {author} {\bibinfo {author} {\bibfnamefont {E.~Z.}\ \bibnamefont {Zhang}}, \bibinfo {author} {\bibfnamefont {L.~E.}\ \bibnamefont {Chern}},\ and\ \bibinfo {author} {\bibfnamefont {Y.~B.}\ \bibnamefont {Kim}},\ }\bibfield  {title} {\bibinfo {title} {Topological magnons for thermal hall transport in frustrated magnets with bond-dependent interactions},\ }\href {https://doi.org/10.1103/PhysRevB.103.174402} {\bibfield  {journal} {\bibinfo  {journal} {Phys. Rev. B}\ }\textbf {\bibinfo {volume} {103}},\ \bibinfo {pages} {174402} (\bibinfo {year} {2021}{\natexlab{b}})}\BibitemShut {NoStop}%
\bibitem [{\citenamefont {Aguilera}\ \emph {et~al.}(2020)\citenamefont {Aguilera}, \citenamefont {Jaeschke-Ubiergo}, \citenamefont {Vidal-Silva}, \citenamefont {Torres},\ and\ \citenamefont {Nunez}}]{Aguilera2020PhysRevB.102.024409}%
  \BibitemOpen
  \bibfield  {author} {\bibinfo {author} {\bibfnamefont {E.}~\bibnamefont {Aguilera}}, \bibinfo {author} {\bibfnamefont {R.}~\bibnamefont {Jaeschke-Ubiergo}}, \bibinfo {author} {\bibfnamefont {N.}~\bibnamefont {Vidal-Silva}}, \bibinfo {author} {\bibfnamefont {L.~E. F.~F.}\ \bibnamefont {Torres}},\ and\ \bibinfo {author} {\bibfnamefont {A.~S.}\ \bibnamefont {Nunez}},\ }\bibfield  {title} {\bibinfo {title} {Topological magnonics in the two-dimensional van der waals magnet ${\mathrm{cri}}_{3}$},\ }\href {https://doi.org/10.1103/PhysRevB.102.024409} {\bibfield  {journal} {\bibinfo  {journal} {Phys. Rev. B}\ }\textbf {\bibinfo {volume} {102}},\ \bibinfo {pages} {024409} (\bibinfo {year} {2020})}\BibitemShut {NoStop}%
\bibitem [{\citenamefont {McClarty}\ \emph {et~al.}(2018)\citenamefont {McClarty}, \citenamefont {Dong}, \citenamefont {Gohlke}, \citenamefont {Rau}, \citenamefont {Pollmann}, \citenamefont {Moessner},\ and\ \citenamefont {Penc}}]{McClarty2018PhysRevB.98.060404}%
  \BibitemOpen
  \bibfield  {author} {\bibinfo {author} {\bibfnamefont {P.~A.}\ \bibnamefont {McClarty}}, \bibinfo {author} {\bibfnamefont {X.-Y.}\ \bibnamefont {Dong}}, \bibinfo {author} {\bibfnamefont {M.}~\bibnamefont {Gohlke}}, \bibinfo {author} {\bibfnamefont {J.~G.}\ \bibnamefont {Rau}}, \bibinfo {author} {\bibfnamefont {F.}~\bibnamefont {Pollmann}}, \bibinfo {author} {\bibfnamefont {R.}~\bibnamefont {Moessner}},\ and\ \bibinfo {author} {\bibfnamefont {K.}~\bibnamefont {Penc}},\ }\bibfield  {title} {\bibinfo {title} {Topological magnons in kitaev magnets at high fields},\ }\href {https://doi.org/10.1103/PhysRevB.98.060404} {\bibfield  {journal} {\bibinfo  {journal} {Phys. Rev. B}\ }\textbf {\bibinfo {volume} {98}},\ \bibinfo {pages} {060404} (\bibinfo {year} {2018})}\BibitemShut {NoStop}%
\bibitem [{\citenamefont {Colpa}(1978)}]{Colpa1978PhysicaA.93.327}%
  \BibitemOpen
  \bibfield  {author} {\bibinfo {author} {\bibfnamefont {J.}~\bibnamefont {Colpa}},\ }\bibfield  {title} {\bibinfo {title} {Diagonalization of the quadratic boson hamiltonian},\ }\href {https://doi.org/https://doi.org/10.1016/0378-4371(78)90160-7} {\bibfield  {journal} {\bibinfo  {journal} {Physica A}\ }\textbf {\bibinfo {volume} {93}},\ \bibinfo {pages} {327} (\bibinfo {year} {1978})}\BibitemShut {NoStop}%
\bibitem [{\citenamefont {Shindou}\ \emph {et~al.}(2013)\citenamefont {Shindou}, \citenamefont {Matsumoto}, \citenamefont {Murakami},\ and\ \citenamefont {Ohe}}]{Shindou2013PhysRevB.87.174427}%
  \BibitemOpen
  \bibfield  {author} {\bibinfo {author} {\bibfnamefont {R.}~\bibnamefont {Shindou}}, \bibinfo {author} {\bibfnamefont {R.}~\bibnamefont {Matsumoto}}, \bibinfo {author} {\bibfnamefont {S.}~\bibnamefont {Murakami}},\ and\ \bibinfo {author} {\bibfnamefont {J.-i.}\ \bibnamefont {Ohe}},\ }\bibfield  {title} {\bibinfo {title} {Topological chiral magnonic edge mode in a magnonic crystal},\ }\href {https://doi.org/10.1103/PhysRevB.87.174427} {\bibfield  {journal} {\bibinfo  {journal} {Phys. Rev. B}\ }\textbf {\bibinfo {volume} {87}},\ \bibinfo {pages} {174427} (\bibinfo {year} {2013})}\BibitemShut {NoStop}%
\bibitem [{\citenamefont {Kim}\ and\ \citenamefont {Kim}(2022)}]{Kim2022PhysRevB.106.104430}%
  \BibitemOpen
  \bibfield  {author} {\bibinfo {author} {\bibfnamefont {H.}~\bibnamefont {Kim}}\ and\ \bibinfo {author} {\bibfnamefont {S.~K.}\ \bibnamefont {Kim}},\ }\bibfield  {title} {\bibinfo {title} {Topological phase transition in magnon bands in a honeycomb ferromagnet driven by sublattice symmetry breaking},\ }\href {https://doi.org/10.1103/PhysRevB.106.104430} {\bibfield  {journal} {\bibinfo  {journal} {Phys. Rev. B}\ }\textbf {\bibinfo {volume} {106}},\ \bibinfo {pages} {104430} (\bibinfo {year} {2022})}\BibitemShut {NoStop}%
\bibitem [{\citenamefont {Zhu}\ \emph {et~al.}(2023)\citenamefont {Zhu}, \citenamefont {Shi}, \citenamefont {Tang},\ and\ \citenamefont {Tang}}]{Zhu2023EurPhysJPlus.138.1045}%
  \BibitemOpen
  \bibfield  {author} {\bibinfo {author} {\bibfnamefont {H.}~\bibnamefont {Zhu}}, \bibinfo {author} {\bibfnamefont {H.}~\bibnamefont {Shi}}, \bibinfo {author} {\bibfnamefont {Z.}~\bibnamefont {Tang}},\ and\ \bibinfo {author} {\bibfnamefont {B.}~\bibnamefont {Tang}},\ }\bibfield  {title} {\bibinfo {title} {Topological phase transitions in a honeycomb ferromagnet with unequal dzyaloshinskii-moriya interactions},\ }\href {https://doi.org/10.1140/epjp/s13360-023-04695-7} {\bibfield  {journal} {\bibinfo  {journal} {Eur. Phys. J. Plus}\ }\textbf {\bibinfo {volume} {138}},\ \bibinfo {pages} {1045} (\bibinfo {year} {2023})}\BibitemShut {NoStop}%
\bibitem [{\citenamefont {R\"uckriegel}\ \emph {et~al.}(2018)\citenamefont {R\"uckriegel}, \citenamefont {Brataas},\ and\ \citenamefont {Duine}}]{Ruckriegel2018PhysRevB.97.081106}%
  \BibitemOpen
  \bibfield  {author} {\bibinfo {author} {\bibfnamefont {A.}~\bibnamefont {R\"uckriegel}}, \bibinfo {author} {\bibfnamefont {A.}~\bibnamefont {Brataas}},\ and\ \bibinfo {author} {\bibfnamefont {R.~A.}\ \bibnamefont {Duine}},\ }\bibfield  {title} {\bibinfo {title} {Bulk and edge spin transport in topological magnon insulators},\ }\href {https://doi.org/10.1103/PhysRevB.97.081106} {\bibfield  {journal} {\bibinfo  {journal} {Phys. Rev. B}\ }\textbf {\bibinfo {volume} {97}},\ \bibinfo {pages} {081106} (\bibinfo {year} {2018})}\BibitemShut {NoStop}%
\bibitem [{\citenamefont {Joshi}(2018)}]{Joshi2018PhysRevB.98.060405}%
  \BibitemOpen
  \bibfield  {author} {\bibinfo {author} {\bibfnamefont {D.~G.}\ \bibnamefont {Joshi}},\ }\bibfield  {title} {\bibinfo {title} {Topological excitations in the ferromagnetic kitaev-heisenberg model},\ }\href {https://doi.org/10.1103/PhysRevB.98.060405} {\bibfield  {journal} {\bibinfo  {journal} {Phys. Rev. B}\ }\textbf {\bibinfo {volume} {98}},\ \bibinfo {pages} {060405} (\bibinfo {year} {2018})}\BibitemShut {NoStop}%
\bibitem [{\citenamefont {Sahu}\ \emph {et~al.}(2021)\citenamefont {Sahu}, \citenamefont {Bhowal},\ and\ \citenamefont {Satpathy}}]{Sahu2021PhysRevB.103.085113}%
  \BibitemOpen
  \bibfield  {author} {\bibinfo {author} {\bibfnamefont {P.}~\bibnamefont {Sahu}}, \bibinfo {author} {\bibfnamefont {S.}~\bibnamefont {Bhowal}},\ and\ \bibinfo {author} {\bibfnamefont {S.}~\bibnamefont {Satpathy}},\ }\bibfield  {title} {\bibinfo {title} {Effect of the inversion symmetry breaking on the orbital hall effect: A model study},\ }\href {https://doi.org/10.1103/PhysRevB.103.085113} {\bibfield  {journal} {\bibinfo  {journal} {Phys. Rev. B}\ }\textbf {\bibinfo {volume} {103}},\ \bibinfo {pages} {085113} (\bibinfo {year} {2021})}\BibitemShut {NoStop}%
\bibitem [{\citenamefont {Pezo}\ \emph {et~al.}(2022)\citenamefont {Pezo}, \citenamefont {Garc\'{\i}a~Ovalle},\ and\ \citenamefont {Manchon}}]{Pezo2022PhysRevB.106.104414}%
  \BibitemOpen
  \bibfield  {author} {\bibinfo {author} {\bibfnamefont {A.}~\bibnamefont {Pezo}}, \bibinfo {author} {\bibfnamefont {D.}~\bibnamefont {Garc\'{\i}a~Ovalle}},\ and\ \bibinfo {author} {\bibfnamefont {A.}~\bibnamefont {Manchon}},\ }\bibfield  {title} {\bibinfo {title} {Orbital hall effect in crystals: Interatomic versus intra-atomic contributions},\ }\href {https://doi.org/10.1103/PhysRevB.106.104414} {\bibfield  {journal} {\bibinfo  {journal} {Phys. Rev. B}\ }\textbf {\bibinfo {volume} {106}},\ \bibinfo {pages} {104414} (\bibinfo {year} {2022})}\BibitemShut {NoStop}%
\bibitem [{\citenamefont {Luttinger}(1964)}]{Luttinger1964PhysRev.135.A1505}%
  \BibitemOpen
  \bibfield  {author} {\bibinfo {author} {\bibfnamefont {J.~M.}\ \bibnamefont {Luttinger}},\ }\bibfield  {title} {\bibinfo {title} {Theory of thermal transport coefficients},\ }\href {https://doi.org/10.1103/PhysRev.135.A1505} {\bibfield  {journal} {\bibinfo  {journal} {Phys. Rev.}\ }\textbf {\bibinfo {volume} {135}},\ \bibinfo {pages} {A1505} (\bibinfo {year} {1964})}\BibitemShut {NoStop}%
\bibitem [{\citenamefont {Matsumoto}\ \emph {et~al.}(2014)\citenamefont {Matsumoto}, \citenamefont {Shindou},\ and\ \citenamefont {Murakami}}]{Matsumoto2014PhysRevB.89.054420}%
  \BibitemOpen
  \bibfield  {author} {\bibinfo {author} {\bibfnamefont {R.}~\bibnamefont {Matsumoto}}, \bibinfo {author} {\bibfnamefont {R.}~\bibnamefont {Shindou}},\ and\ \bibinfo {author} {\bibfnamefont {S.}~\bibnamefont {Murakami}},\ }\bibfield  {title} {\bibinfo {title} {Thermal hall effect of magnons in magnets with dipolar interaction},\ }\href {https://doi.org/10.1103/PhysRevB.89.054420} {\bibfield  {journal} {\bibinfo  {journal} {Phys. Rev. B}\ }\textbf {\bibinfo {volume} {89}},\ \bibinfo {pages} {054420} (\bibinfo {year} {2014})}\BibitemShut {NoStop}%
\bibitem [{\citenamefont {Zyuzin}\ and\ \citenamefont {Kovalev}(2016)}]{Zyuzin2016PhysRevLett.117.217203}%
  \BibitemOpen
  \bibfield  {author} {\bibinfo {author} {\bibfnamefont {V.~A.}\ \bibnamefont {Zyuzin}}\ and\ \bibinfo {author} {\bibfnamefont {A.~A.}\ \bibnamefont {Kovalev}},\ }\bibfield  {title} {\bibinfo {title} {Magnon spin nernst effect in antiferromagnets},\ }\href {https://doi.org/10.1103/PhysRevLett.117.217203} {\bibfield  {journal} {\bibinfo  {journal} {Phys. Rev. Lett.}\ }\textbf {\bibinfo {volume} {117}},\ \bibinfo {pages} {217203} (\bibinfo {year} {2016})}\BibitemShut {NoStop}%
\bibitem [{\citenamefont {Li}\ \emph {et~al.}(2020{\natexlab{b}})\citenamefont {Li}, \citenamefont {Sandhoefner},\ and\ \citenamefont {Kovalev}}]{Li2020PhysRevRes.2.013079}%
  \BibitemOpen
  \bibfield  {author} {\bibinfo {author} {\bibfnamefont {B.}~\bibnamefont {Li}}, \bibinfo {author} {\bibfnamefont {S.}~\bibnamefont {Sandhoefner}},\ and\ \bibinfo {author} {\bibfnamefont {A.~A.}\ \bibnamefont {Kovalev}},\ }\bibfield  {title} {\bibinfo {title} {Intrinsic spin nernst effect of magnons in a noncollinear antiferromagnet},\ }\href {https://doi.org/10.1103/PhysRevResearch.2.013079} {\bibfield  {journal} {\bibinfo  {journal} {Phys. Rev. Res.}\ }\textbf {\bibinfo {volume} {2}},\ \bibinfo {pages} {013079} (\bibinfo {year} {2020}{\natexlab{b}})}\BibitemShut {NoStop}%
\bibitem [{\citenamefont {Owerre}(2016)}]{Owerre2016JApplPhys.120.043903}%
  \BibitemOpen
  \bibfield  {author} {\bibinfo {author} {\bibfnamefont {S.~A.}\ \bibnamefont {Owerre}},\ }\bibfield  {title} {\bibinfo {title} {{Topological honeycomb magnon Hall effect: A calculation of thermal Hall conductivity of magnetic spin excitations}},\ }\href {https://doi.org/10.1063/1.4959815} {\bibfield  {journal} {\bibinfo  {journal} {J. Appl. Phys.}\ }\textbf {\bibinfo {volume} {120}},\ \bibinfo {pages} {043903} (\bibinfo {year} {2016})}\BibitemShut {NoStop}%
\bibitem [{\citenamefont {Dyrda\l{}}\ \emph {et~al.}(2016{\natexlab{a}})\citenamefont {Dyrda\l{}}, \citenamefont {Barna\ifmmode~\acute{s}\else \'{s}\fi{}},\ and\ \citenamefont {Dugaev}}]{Dyrdal2016PhysRevB.94.035306}%
  \BibitemOpen
  \bibfield  {author} {\bibinfo {author} {\bibfnamefont {A.}~\bibnamefont {Dyrda\l{}}}, \bibinfo {author} {\bibfnamefont {J.}~\bibnamefont {Barna\ifmmode~\acute{s}\else \'{s}\fi{}}},\ and\ \bibinfo {author} {\bibfnamefont {V.~K.}\ \bibnamefont {Dugaev}},\ }\bibfield  {title} {\bibinfo {title} {Spin hall and spin nernst effects in a two-dimensional electron gas with rashba spin-orbit interaction: Temperature dependence},\ }\href {https://doi.org/10.1103/PhysRevB.94.035306} {\bibfield  {journal} {\bibinfo  {journal} {Phys. Rev. B}\ }\textbf {\bibinfo {volume} {94}},\ \bibinfo {pages} {035306} (\bibinfo {year} {2016}{\natexlab{a}})}\BibitemShut {NoStop}%
\bibitem [{\citenamefont {Dyrda\l{}}\ \emph {et~al.}(2016{\natexlab{b}})\citenamefont {Dyrda\l{}}, \citenamefont {Dugaev},\ and\ \citenamefont {Barna\ifmmode~\acute{s}\else \'{s}\fi{}}}]{Dyrdal2016PhysRevB.94.205302}%
  \BibitemOpen
  \bibfield  {author} {\bibinfo {author} {\bibfnamefont {A.}~\bibnamefont {Dyrda\l{}}}, \bibinfo {author} {\bibfnamefont {V.~K.}\ \bibnamefont {Dugaev}},\ and\ \bibinfo {author} {\bibfnamefont {J.}~\bibnamefont {Barna\ifmmode~\acute{s}\else \'{s}\fi{}}},\ }\bibfield  {title} {\bibinfo {title} {Spin-resolved orbital magnetization in rashba two-dimensional electron gas},\ }\href {https://doi.org/10.1103/PhysRevB.94.205302} {\bibfield  {journal} {\bibinfo  {journal} {Phys. Rev. B}\ }\textbf {\bibinfo {volume} {94}},\ \bibinfo {pages} {205302} (\bibinfo {year} {2016}{\natexlab{b}})}\BibitemShut {NoStop}%
\bibitem [{\citenamefont {McGuire}\ \emph {et~al.}(2015)\citenamefont {McGuire}, \citenamefont {Dixit}, \citenamefont {Cooper},\ and\ \citenamefont {Sales}}]{McGuire2015ChemMater.27.612}%
  \BibitemOpen
  \bibfield  {author} {\bibinfo {author} {\bibfnamefont {M.~A.}\ \bibnamefont {McGuire}}, \bibinfo {author} {\bibfnamefont {H.}~\bibnamefont {Dixit}}, \bibinfo {author} {\bibfnamefont {V.~R.}\ \bibnamefont {Cooper}},\ and\ \bibinfo {author} {\bibfnamefont {B.~C.}\ \bibnamefont {Sales}},\ }\bibfield  {title} {\bibinfo {title} {Coupling of crystal structure and magnetism in the layered, ferromagnetic insulator $\mathrm{CrI}_{3}$},\ }\href {https://doi.org/10.1021/cm504242t} {\bibfield  {journal} {\bibinfo  {journal} {Chem. Mater.}\ }\textbf {\bibinfo {volume} {27}},\ \bibinfo {pages} {612} (\bibinfo {year} {2015})}\BibitemShut {NoStop}%
\bibitem [{\citenamefont {Janssen}\ \emph {et~al.}(2017)\citenamefont {Janssen}, \citenamefont {Andrade},\ and\ \citenamefont {Vojta}}]{Janssen2017PhysRevB.96.064430}%
  \BibitemOpen
  \bibfield  {author} {\bibinfo {author} {\bibfnamefont {L.}~\bibnamefont {Janssen}}, \bibinfo {author} {\bibfnamefont {E.~C.}\ \bibnamefont {Andrade}},\ and\ \bibinfo {author} {\bibfnamefont {M.}~\bibnamefont {Vojta}},\ }\bibfield  {title} {\bibinfo {title} {Magnetization processes of zigzag states on the honeycomb lattice: Identifying spin models for $\ensuremath{\alpha}\text{\ensuremath{-}}{\mathrm{rucl}}_{3}$ and ${\mathrm{na}}_{2}{\mathrm{iro}}_{3}$},\ }\href {https://doi.org/10.1103/PhysRevB.96.064430} {\bibfield  {journal} {\bibinfo  {journal} {Phys. Rev. B}\ }\textbf {\bibinfo {volume} {96}},\ \bibinfo {pages} {064430} (\bibinfo {year} {2017})}\BibitemShut {NoStop}%
\bibitem [{\citenamefont {Chern}\ \emph {et~al.}(2020)\citenamefont {Chern}, \citenamefont {Kaneko}, \citenamefont {Lee},\ and\ \citenamefont {Kim}}]{Li2020PhysRevRes.2.013014}%
  \BibitemOpen
  \bibfield  {author} {\bibinfo {author} {\bibfnamefont {L.~E.}\ \bibnamefont {Chern}}, \bibinfo {author} {\bibfnamefont {R.}~\bibnamefont {Kaneko}}, \bibinfo {author} {\bibfnamefont {H.-Y.}\ \bibnamefont {Lee}},\ and\ \bibinfo {author} {\bibfnamefont {Y.~B.}\ \bibnamefont {Kim}},\ }\bibfield  {title} {\bibinfo {title} {Magnetic field induced competing phases in spin-orbital entangled kitaev magnets},\ }\href {https://doi.org/10.1103/PhysRevResearch.2.013014} {\bibfield  {journal} {\bibinfo  {journal} {Phys. Rev. Res.}\ }\textbf {\bibinfo {volume} {2}},\ \bibinfo {pages} {013014} (\bibinfo {year} {2020})}\BibitemShut {NoStop}%
\bibitem [{\citenamefont {Ibañez-Azpiroz}\ \emph {et~al.}(2022)\citenamefont {Ibañez-Azpiroz}, \citenamefont {de~Juan},\ and\ \citenamefont {Souza}}]{IbanezAzpiroz2022SciPostPhys.12.070}%
  \BibitemOpen
  \bibfield  {author} {\bibinfo {author} {\bibfnamefont {J.}~\bibnamefont {Ibañez-Azpiroz}}, \bibinfo {author} {\bibfnamefont {F.}~\bibnamefont {de~Juan}},\ and\ \bibinfo {author} {\bibfnamefont {I.}~\bibnamefont {Souza}},\ }\bibfield  {title} {\bibinfo {title} {{Assessing the role of interatomic position matrix elements in tight-binding calculations of optical properties}},\ }\href {https://doi.org/10.21468/SciPostPhys.12.2.070} {\bibfield  {journal} {\bibinfo  {journal} {SciPost Phys.}\ }\textbf {\bibinfo {volume} {12}},\ \bibinfo {pages} {070} (\bibinfo {year} {2022})}\BibitemShut {NoStop}%
\bibitem [{\citenamefont {Thonhauser}(2011)}]{Thonhauser2011IntJModPhysB25.1429}%
  \BibitemOpen
  \bibfield  {author} {\bibinfo {author} {\bibfnamefont {T.}~\bibnamefont {Thonhauser}},\ }\bibfield  {title} {\bibinfo {title} {Theory of orbital magnetization in solids},\ }\href {https://doi.org/10.1142/S0217979211058912} {\bibfield  {journal} {\bibinfo  {journal} {Int. J. Mod. Phys. B}\ }\textbf {\bibinfo {volume} {25}},\ \bibinfo {pages} {1429} (\bibinfo {year} {2011})}\BibitemShut {NoStop}%
\bibitem [{\citenamefont {Bhowal}\ and\ \citenamefont {Vignale}(2021)}]{Bhowal2021PhysRevB.103.195309}%
  \BibitemOpen
  \bibfield  {author} {\bibinfo {author} {\bibfnamefont {S.}~\bibnamefont {Bhowal}}\ and\ \bibinfo {author} {\bibfnamefont {G.}~\bibnamefont {Vignale}},\ }\bibfield  {title} {\bibinfo {title} {Orbital hall effect as an alternative to valley hall effect in gapped graphene},\ }\href {https://doi.org/10.1103/PhysRevB.103.195309} {\bibfield  {journal} {\bibinfo  {journal} {Phys. Rev. B}\ }\textbf {\bibinfo {volume} {103}},\ \bibinfo {pages} {195309} (\bibinfo {year} {2021})}\BibitemShut {NoStop}%
\end{thebibliography}%

\clearpage
\begin{appendix}
\begin{widetext}
\renewcommand\thefigure{\thesection\arabic{figure}}

\section{Quadratic bosonic Hamiltonian}\label{Appendix A}
\setcounter{figure}{0}

To understand our total Hamiltonian, we need three bases; the system coordinate basis $\{\bdsym{x},\bdsym{y},\bdsym{z}\}$ describes our system as a honeycomb structure and $z$-axis is parallel to the DM interaction vector, the Kitaev spin coordinate basis $\{\bdsym{a},\bdsym{b},\bdsym{c}\}$ is the orientation of the spins which couples through the Kitaev interaction, and the spin-polarized coordinate basis $\{\bdsym{\xi},\bdsym{\zeta},\bdsym{n}\}$ is the spin basis of the Holstein-Primakoff transform.
To describe the linear spin-wave magnon, we need to write all interactions consisting of the Hamiltonian in the spin-polarized basis using the rotation matrix $R(\vartheta,\varphi)$:
\begin{equation}
    R(\vartheta,\varphi)=
    \begin{bmatrix}
        \cos{\vartheta}\cos{\varphi} & -\sin{\varphi} & \sin{\vartheta}\cos{\varphi}\\
        \cos{\vartheta}\sin{\varphi} & \cos{\varphi} & \sin{\vartheta}\sin{\varphi}\\
        -\sin{\vartheta} & 0 & \cos{\vartheta}
    \end{bmatrix}.
\end{equation}
We transform $( \hat{S}^{x}_i, \hat{S}^{y}_i, \hat{S}^{z}_i)^{T}=R(\vartheta,\varphi)( \hat{S}^{\xi}_i, \hat{S}^{\zeta}_i, \hat{S}^{n}_i)^{T}$ for the DM and Zeeman interaction, and $( \hat{S}^{a}_i, \hat{S}^{b}_i, \hat{S}^{c}_i)^{T}=R_{\text{K}}( \hat{S}^{x}_i, \hat{S}^{y}_i, \hat{S}^{z}_i)^{T}=R_{\text{K}}R(\vartheta,\varphi)( \hat{S}^{\xi}_i, \hat{S}^{\zeta}_i, \hat{S}^{n}_i)^{T}$ for the Kitaev interaction, where $R_{\text{K}}=[\bdsym{a},\bdsym{b},\bdsym{c}]^{T}$ where $R_{\text{K}}$ is a sample specific 3-by-3 matrix of the column vectors $\bdsym{a}$, $\bdsym{b}$, $\bdsym{c}$ in the system coordinate.
Each interaction Hamiltonian is written down by

\begin{equation}\label{Hquad}
    \begin{split}
        &H = H_{\text{ex}}+H_{\text{K}}+H_{\text{DM}}+H_{\text{Z}},\\
        &H_{\text{ex}} =-JS\sum_{\langle i,j\rangle}[S-(\hat{a}_i^{\dagger} \hat{a}_i+\hat{b}_j^{\dagger}\hat{b}_j)+\hat{a}_i \hat{b}_j^{\dagger}+\hat{a}_i^{\dagger} \hat{b}_j],\\
        &H_{\text{DM}} =-iDS\Bigg[\sum_{\langle\langle  i,j\rangle\rangle}\nu_{ij}\left(\hat{a}_{i}\hat{a}^{\dagger}_{j}-\hat{a}^{\dagger}_{i}\hat{a}_{j}\right)\cos{\theta}
        +\sum_{\langle\langle  i,j\rangle\rangle}\nu_{ij}\left(\hat{b}_{i}\hat{b}^{\dagger}_{j}-\hat{b}^{\dagger}_{i}\hat{b}_{j}\right)\cos{\theta}\Bigg],\\
        &H_{\text{Z}}  =-g\mu_B B\sum_{i}\Bigg[
        (S-\hat{\alpha}^{\dagger}_{i} \hat{\alpha}_{i})(\sin\theta\sin\vartheta\cos(\varphi-\phi)+\cos\theta\cos\vartheta)\\
        &+\sqrt{\frac{S}{2}}\Bigg((\hat{\alpha}^{\dagger}_i+\hat{\alpha}_i)
        (\sin\theta\cos\vartheta\cos(\varphi-\phi)-\sin\vartheta\cos\theta)
        -i(\hat{\alpha}^{\dagger}_i-\hat{\alpha}_i)
        \sin\theta\sin(\varphi-\phi)\Bigg)
        \Bigg],
    \end{split}
\end{equation}
where $\hat{\alpha}^{\dagger}_i$ and $\hat{\alpha}_i$ are the creation and annihilation operator at $i$ and if $i$ is in $A$ ($B$) sublattice $\alpha$ is identical to $\hat{a}$ ($\hat{b}$), and
\begin{equation}\label{H_Kit}
    \begin{split}
        &H_{\text{K}} 
        =-\sum_{\langle i,j\rangle_{\gamma}} \Bigg(
        S[H_{\text{K}}^{\gamma,n}]_{33}\Big(
        S-\hat{a}^{\dagger}_i \hat{a}_i-\hat{b}^{\dagger}_j \hat{b}_j
        \Big)
        +S\sqrt{\frac{S}{2}}\Big[
        ([H_{\text{K}}^{\gamma,n}]_{13}+i[H_{\text{K}}^{\gamma,n}]_{23})(\hat{a}^{\dagger}_i+\hat{b}^{\dagger}_j)+([H_{\text{K}}^{\gamma,n}]_{13}-i[H_{\text{K}}^{\gamma,n}]_{23})(\hat{a}_i+\hat{b}_j)
        \Big]\\
        &+\frac{S}{2}
        \begin{bmatrix}
            \hat{a}^{\dagger}_i& \hat{a}_i
        \end{bmatrix}
        \begin{bmatrix}
            ([H_{\text{K}}^{\gamma,n}]_{11}-[H_{\text{K}}^{\gamma,n}]_{22})+i([H_{\text{K}}^{\gamma,n}]_{12}+[H_{\text{K}}^{\gamma,n}]_{21}) &
            ([H_{\text{K}}^{\gamma,n}]_{11}+[H_{\text{K}}^{\gamma,n}]_{22})\\
            ([H_{\text{K}}^{\gamma,n}]_{11}+[H_{\text{K}}^{\gamma,n}]_{22})&
            ([H_{\text{K}}^{\gamma,n}]_{11}-[H_{\text{K}}^{\gamma,n}]_{22})-i([H_{\text{K}}^{\gamma,n}]_{12}+[H_{\text{K}}^{\gamma,n}]_{21})
        \end{bmatrix}
        \begin{bmatrix}
            \hat{b}^{\dagger}_j \\ \hat{b}_j
        \end{bmatrix}
        \Bigg),
    \end{split}
\end{equation}
where $[H_{\text{K}}^{\gamma,n}]_{ij}$ is the $i,j$ component of the matrix form of $H_{\text{K}}^{\gamma,n}$ in the spin-polarized coordinate basis, which is obtained as follows.

The Kitaev Hamiltonian, $H_{\text{K}} = -\sum_{\langle i,j\rangle_{\gamma}}  H_{\text{K}}^{\langle i,j\rangle_{\gamma}}$, of coupling between site $i$ and $j$ with bonding type $\gamma$ is written by 
\begin{equation}
    \begin{split}
        H_{\text{K}}^{\langle i,j\rangle_{\gamma}} 
        =&
        \begin{bmatrix}
             \hat{S}^{a}_i &  \hat{S}^{b}_i &  \hat{S}^{c}_i
        \end{bmatrix}
        H_{\text{K}}^{\gamma}
        \begin{bmatrix}
             \hat{S}^{a}_j\\ \hat{S}^{b}_j\\ \hat{S}^{c}_j
        \end{bmatrix}
    \end{split},
\end{equation}
where
\begin{equation}
    \begin{split}
    H_{\text{K}}^{\gamma=a}=
        \begin{bmatrix}
            K & \Gamma' & \Gamma'\\
            \Gamma' & 0 & \Gamma\\
            \Gamma' & \Gamma & 0
        \end{bmatrix},&\quad
    H_{\text{K}}^{\gamma=b}=
        \begin{bmatrix}
            0 & \Gamma' & \Gamma\\
            \Gamma' & K & \Gamma'\\
            \Gamma & \Gamma' & 0
            \end{bmatrix},\quad
    H_{\text{K}}^{\gamma=c}=
        \begin{bmatrix}
            0 & \Gamma & \Gamma'\\
            \Gamma & 0 & \Gamma'\\
            \Gamma' & \Gamma' & K
        \end{bmatrix}.
    \end{split}
\end{equation}
In the spin-polarized coordinate, the Kitaev Hamiltonian is
\begin{equation}
    \begin{split}
        H_{\text{K}}^{\langle i,j\rangle_{\gamma}} 
        =&
        \begin{bmatrix}
             \hat{S}^{\xi}_i &  \hat{S}^{\zeta}_i &  \hat{S}^{n}_i
        \end{bmatrix}
        H_{\text{K}}^{\gamma,n}
        \begin{bmatrix}
             \hat{S}^{\xi}_j\\ \hat{S}^{\zeta}_j\\ \hat{S}^{n}_j
        \end{bmatrix}
    \end{split},
\end{equation}
and we define $H_{\text{K}}^{\gamma,n}=R^{T}(\vartheta,\varphi)R_{\text{K}}^{T}H_{\text{K}}^{\gamma}R_{\text{K}}R(\vartheta,\varphi)$.

The spin polarization angles $\varphi$ and $\vartheta$ are determined to remove the sum of the term in the second line of the Zeeman interaction in Eq.~\eqref{Hquad} and the second term enclosed by the square bracket in Eq.~\eqref{H_Kit}.
If such $\varphi$ and $\vartheta$ do not exist, i.e., there is the linear order of the bosonic operators, the fully spin-polarized phase is not the ground state, thus we need to consider other phases like zig-zag phases~\cite{Zhang2021PhysRevB.103.174402}. 

For our system, that is identical to $\mathrm{CrI}_{3}$ case, with that Kitaev vectors are orthogonal to each other, the Kitaev vectors are $\bdsym{a}=(-1/\sqrt{6},1/\sqrt{2},1/\sqrt{3})^{T}$, $\bdsym{b}=(-1/\sqrt{6},-1/\sqrt{2},1/\sqrt{3})^{T}$, and $\bdsym{c}=(\sqrt{2/3},0,1/\sqrt{3})^{T}$~\cite{Aguilera2020PhysRevB.102.024409,Zhang2021PhysRevB.103.134414}, so that
\begin{equation}\label{Cry2Coord}
    R_{\text{K}}=
    \begin{bmatrix}
        -1/\sqrt{6} & 1/\sqrt{2} & 1/\sqrt{3}\\
        -1/\sqrt{6} & -1/\sqrt{2} & 1/\sqrt{3}\\
        \sqrt{2/3} & 0 & 1/\sqrt{3}
    \end{bmatrix}.
\end{equation}
The condition for the $\varphi$ and $\vartheta$ is the one minimizing the energy among $\varphi=\phi$ or $\phi+\pi$ and $\vartheta$ satisfying $\pm\sin(\theta\mp \vartheta)/\sin2\vartheta=3S(\Gamma+2\Gamma')/2g\mu_B B$ which condition with $\Gamma'=0$ reduces to equation (12) in the paper~\cite{Janssen2017PhysRevB.96.064430}.
For the normal magnetic field configuration $\bdsym{B}=B\bdsym{z}$, which our system considers, the spins are in the fully polarized phase along the external magnetic field~\cite{Li2020PhysRevRes.2.013014}, i.e., $\vartheta=0$.

\section{Hamiltonian matrix and magnon energy band}\label{Appendix B}
\setcounter{figure}{0}

By collecting the Hamiltonian in Eq.~(\ref{Hquad}) and Eq.~(\ref{H_Kit}) and performing the Fourier transform, we obtain
\begin{equation}
    H = C
    +\frac{1}{2}\sum_{\bdsym{k}}\hat{\Psi}_{\bdsym{k}}^{\dagger}H_{\bdsym{k}}\hat{\Psi}_{\bdsym{k}},\quad
    H_{\bdsym{k}} =
    \begin{bmatrix}
        B^d_{\bdsym{k}} & B^o_{\bdsym{k}}\\
        (B^o_{\bdsym{k}})^{\dagger} & (B^d_{\bdsym{-k}})^*
    \end{bmatrix}
\end{equation}
where $\hat{\Psi}_{\bdsym{k}}=(\hat{a}_{\bdsym{k}},\hat{b}_{\bdsym{k}},\hat{a}^{\dagger}_{-\bdsym{k}},\hat{b}^{\dagger}_{-\bdsym{k}})$, $^{\dagger}$ is hermitian conjugate, and $^*$ is complex conjugate,
\begin{equation}\label{Hmat}
\begin{split}
    B^d_{\bdsym{k}} &=
    \begin{bmatrix}
        E^{0}(\bdsym{k})-\Delta(\bdsym{k}) &
        T^{+}_r(\bdsym{k})+\Tilde{J}(\bdsym{k})&
        \\
        T^{+}_r(-\bdsym{k})+\Tilde{J}(-\bdsym{k}) &
        E^{0}(\bdsym{k})+\Delta(\bdsym{k}) &
    \end{bmatrix},\\
    B^o_{\bdsym{k}} &=
    \begin{bmatrix}
        0&
        T^{-}_r(\bdsym{k})+iO_{ff}(\bdsym{k})
        \\
        T^{-}_r(-\bdsym{k})+iO_{ff}(-\bdsym{k}) &
        0
    \end{bmatrix},
\end{split}
\end{equation}
where $E^{0}(\bdsym{k})=3JS+g\mu_B B(\sin\theta\sin\vartheta\cos(\varphi-\phi)+\cos\theta\cos\vartheta)+S\sum_{\gamma\in\{a,b,c\}}[H_{\text{K}}^{\gamma,n}]_{33}$ which contains the Zeeman, Heisenberg, Kitaev interactions,
$\Tilde{J}(\bdsym{k})=-\sum_{\gamma\in\{a,b,c\}}JSe^{i\bdsym{k}\cdot \bdsym{d}_{\gamma}}$ which come from the Heisenberg interaction,
$\Delta(\bdsym{k})=2DS\sum_{i}\sin(\bdsym{k}\cdot \bdsym{R}_i)\cos{\vartheta}$ which is resulted from the DM interaction, and
$T^{\pm}_r(\bdsym{k})=-\sum_{\gamma\in\{a,b,c\}}\Big([H_{\text{K}}^{\gamma,n}]_{11}\pm [H_{\text{K}}^{\gamma,n}]_{22}\Big)Se^{i\bdsym{k}\cdot \bdsym{d}_{\gamma}}/2$ and
$O_{ff}(\bdsym{k})=-\sum_{\gamma\in\{a,b,c\}}\Big([H_{\text{K}}^{\gamma,n}]_{12}+[H_{\text{K}}^{\gamma,n}]_{21}\Big)Se^{i\bdsym{k}\cdot \bdsym{d}_{\gamma}}/2$ which are come from the Kitaev interaction and make us perform the Bogoliubov transform due to introducing the non-conserving magnon number term in $\hat{\Psi}_{\bdsym{k}}$ basis.

Diagonalizing the Hamiltonian is done by the Bogoliubov transformation using the Cholesky decomposition~\cite{Colpa1978PhysicaA.93.327,Shindou2013PhysRevB.87.174427}.
In detail, using an upper triangular matrix $C$ from the Cholesky decomposition $H_{\bdsym{k}}=C^{\dagger}C$, diagonalization of $C\eta C^{\dagger}=P^{\dagger}\bar{E}P$, where $\eta=\text{diag}(1,1,-1,-1)$ and $P$ is unitary matrix, gives diagonal matrix $\bar{E}(\bdsym{k})$ which directly connects with $E(\bdsym{k})=\eta\bar{E}(\bdsym{k})$ if $\bar{E}(\bdsym{k})$ is sorted with large to small order.
The corresponding eigenvectors are column vectors of $U=C^{-1}P^{\dagger}E^{1/2}(\bdsym{k})$, which is para-unitary, i.e., $U^{-1}=\eta U^{\dagger}\eta$.
Hence the left eigenvectors $\langle u_n^L(\bdsym{k})|$ and right eigenvectors $|u_n^R(\bdsym{k})\rangle$ of $\bar{H}_{\bdsym{k}}$ are from the row vector of $\eta U^{\dagger}\eta$ and column vector of $U$ in the matrix form with satisfying $\langle u_n^L(\bdsym{k})|u_m^R(\bdsym{k})\rangle =\delta_{nm}$, i.e., $\bar{H}_{\bdsym{k}}U=U\bar{E}(\bdsym{k})$, $\eta U^{\dagger}\eta\bar{H}_{\bdsym{k}}=\bar{E}(\bdsym{k})\eta U^{\dagger}\eta$.
From now on, we use $|u_n(\bdsym{k})\rangle\equiv|u_n^R(\bdsym{k})\rangle$ and $\langle u_n(\bdsym{k})|= (|u_n(\bdsym{k})\rangle)^{\dagger}= \eta_{nn}\langle u_n^L(\bdsym{k})|\eta$.

We can obtain magnon energy band by extracting $E(\bdsym{k})=\eta\bar{E'}(\bdsym{k})$ where $\bar{E'}(\bdsym{k})$ is eigenvalues of $\eta H_{\bdsym{k}}$.
However, this process does not give proper eigenvectors, thus to evaluate Berry curvature and MOMBC we obtain the energy and eigenvector pair by the Cholesky decomposition process.
The energy eigenvalues at $K$ and $K'$ points are
\begin{equation}\label{Kpt}
    \sqrt{(g\mu_B B+3Sz)^2+\left(3 \sqrt{3} D S\right)^2 -2(\frac{KS}{2})^2\pm2\left| \left((\frac{KS}{2})^2+3\sqrt{3} D S (g\mu_B B+ 3Sz) \right)\right|},
\end{equation}
where $z=J+K/3$.
From Eq.~\eqref{Kpt}, gap closing occurs at $3\sqrt{3} DS =-(KS/2)^2/(3Sz+g\mu_B B)$.
That is the line of topological phase transition.

\section{Magnon orbital moment Nernst conductivity}\label{Appendix C}
\setcounter{figure}{0} 
The $n$th band Berry connection $\bdsym{A}_n(\bdsym{k})$ is defined by $\bdsym{A}_n(\bdsym{k})=\eta_{nn}\langle u_n(\bdsym{k})|\eta|i\partial_{\bdsym{k}} u_n(\bdsym{k})\rangle$~\cite{Shindou2013PhysRevB.87.174427,Go2024NanoLett.24.5968}, and the $n$th band Berry curvature is defined by $\Omega_n(\bdsym{k})=\partial_{\bdsym{k}}\times\bdsym{A}_n(\bdsym{k})$,
\begin{equation}\label{BerryCurv}
    \Omega_n(\bdsym{k})=i\eta_{nn}\langle i\partial_{\bdsym{k}}u_n(\bdsym{k})|\times\eta|i\partial_{\bdsym{k}} u_n(\bdsym{k})\rangle\cdot \bdsym{z}.
\end{equation}
Using the identity
\begin{equation}\label{iden}
    |i\bdsym{\partial}_k u_n(\bdsym{k})\rangle
    =\sum_{m\neq n}\eta_{mm}|u_m(\bdsym{k})\rangle\frac{\langle u_m(\bdsym{k})|i\hbar\hat{\bdsym{v}} |u_n(\bdsym{k})\rangle}{\bar{\epsilon}_n(\bdsym{k})-\bar{\epsilon}_m(\bdsym{k})},
\end{equation}
we obtain the $n$th band Berry curvature
\begin{equation}\label{nthBerryCurv}
\Omega_n(\bdsym{k})=i\sum_{m\neq n}\frac{\eta_{nn}\eta_{mm}\hbar^{2}}{(\bar{\epsilon}_n(\bdsym{k})-\bar{\epsilon}_m(\bdsym{k}))^2}
\langle u_n(\bdsym{k})|\hat{\bdsym{v}}
|u_m(\bdsym{k})\rangle
\times \langle u_m(\bdsym{k})|\hat{\bdsym{v}}|u_n(\bdsym{k})\rangle\cdot \bdsym{z},
\end{equation}
here $\bar{\epsilon}_{n}(\bdsym{k})=[\bar{E}(\bdsym{k})]_{nn}$.

The MOMNC~\cite{Go2024NanoLett.24.5968} is 
\begin{equation}
    \alpha^{m_z}_{yx}=\frac{2k_B}{V}\sum_{n}\sum_{\bdsym{k}}c_1(g_n(\bdsym{k}))\Omega^{m_z}_{n}(\bdsym{k}),
\end{equation}
which originated from the statistical weight of the linear response of the magnon orbital moment current density from the temperature gradient, $J^{m_z}_{y} = -\alpha^{m_z}_{yx}\partial_{x}T$, as obtained from the spin case~\cite{Li2020PhysRevRes.2.013079}.

The magnon orbital moment Berry curvature (MOMBC) $\Omega^{m_z}_{n}(\bdsym{k})$ is written in Eq.~(\ref{Lberry2}).
\begin{equation}\label{Lberry2}
\Omega^{m_z}_n(\bdsym{k})=-2\hbar \sum_{m\neq n}\text{Im}\Big[\frac{\eta_{nn}\eta_{mm}}{(\bar{\epsilon}_{n}(\bdsym{k})-\bar{\epsilon}_{m}(\bdsym{k}))^2}
\times\langle u_n(\bdsym{k})|\hat{j}^{m_z}_{y}|u_m(\bdsym{k})\rangle\langle u_m(\bdsym{k})|\hat{v}_x|u_n(\bdsym{k})\rangle\Big],
\end{equation}
To compute this, we need to calculate $\langle u_n(\bdsym{k})|\hat{j}^{m_z}_{y}|u_m(\bdsym{k})\rangle$ and $\langle u_m(\bdsym{k})|\hat{v}_x|u_n(\bdsym{k})\rangle$.
In the right eigenstate basis, the $nm$-component of the magnon orbital moment current is
\begin{equation}\label{jzy}
\begin{split}
    &[j^{m_z}_{y}]_{nm}=\langle u_n(\bdsym{k})|\hat{j}^{m_z}_{y}|u_m(\bdsym{k})\rangle
    =\frac{1}{4}\langle u_n(\bdsym{k})|\hat{m}_{z}\eta\hat{v}_{y}+\hat{v}_{y}\eta\hat{m}_{z}|u_m(\bdsym{k})\rangle\\
    =&\sum_{p}\frac{\eta_{pp}}{4}\Big[\langle u_n(\bdsym{k})|\hat{m}_{z}|u_p(\bdsym{k})\rangle\langle u_p(\bdsym{k})|\hat{v}_{y}|u_m(\bdsym{k})\rangle
    +\langle u_n(\bdsym{k})|\hat{v}_{y}|u_p(\bdsym{k})\rangle\langle u_p(\bdsym{k})|\hat{m}_{z}|u_m(\bdsym{k})\rangle\Big],
\end{split}
\end{equation}
where the $nm$-component of the magnon orbital moment is
\begin{equation}\label{m_z}
[\bdsym{m}]_{nm}=\langle u_n(\bdsym{k})|\hat{\bdsym{m}}|u_m(\bdsym{k})\rangle = \sum_{q\in \{n,m\}}\sum_{p\neq q}\frac{1}{\bar{\epsilon}_{p}(\bdsym{k})-\bar{\epsilon}_{q}(\bdsym{k})}
\times\frac{i\hbar\eta_{pp}}{4}\Big(\langle u_n(\bdsym{k})|\hat{\bdsym{v}}|u_p(\bdsym{k})\rangle\times\langle u_p(\bdsym{k})|\hat{\bdsym{v}}|u_m(\bdsym{k})\rangle\Big).
\end{equation}
The magnon orbital moment is obtained using the position operator~\cite{IbanezAzpiroz2022SciPostPhys.12.070,Pezo2022PhysRevB.106.104414} and perturbation identity of the derivative of $\bdsym{k}$~\cite{Thonhauser2011IntJModPhysB25.1429,Bhowal2021PhysRevB.103.195309} acting on the cell-periodic Bloch function $|u_n(\bdsym{k})\rangle$ with applying Eq.~(\ref{iden}),
\begin{equation}\label{r}
    \hat{\bdsym{r}}|u_n(\bdsym{k})\rangle=|i\bdsym{\partial}_k u_n(\bdsym{k})\rangle.
\end{equation}

\section{Effect of the magnetic field on the energy band, Berry curvature, and magnon orbital moment Berry curvature}\label{Appendix F}
\setcounter{figure}{0}
In this section, we show the ratio between the additional magnon energy increase to Zeeman energy and the magnon energy, and the energy band, Berry curvature, and MOMBC of the $J$-$K$-$B$ model with different external magnetic fields.

\begin{figure}
    \centering
    \includegraphics[width=500pt]{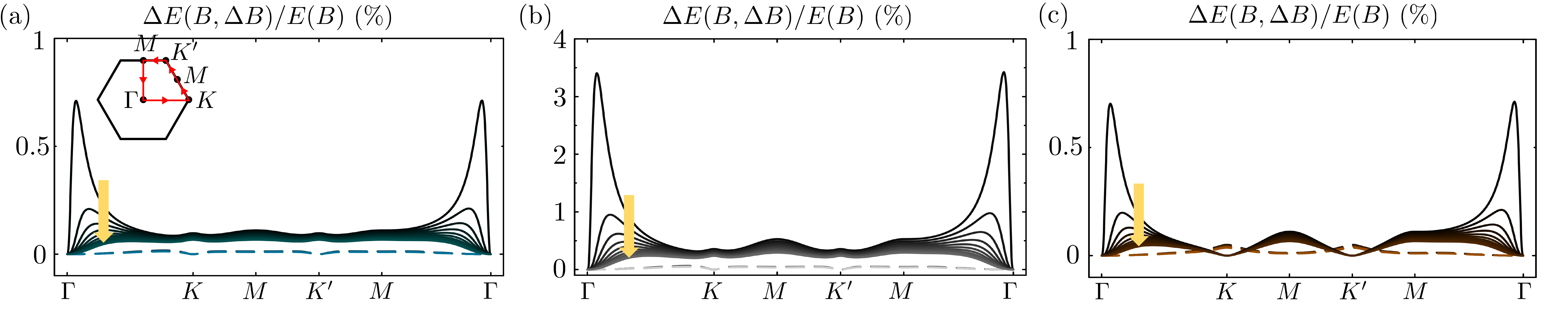} 
    \caption{The ratio between the additional magnon energy increase to Zeeman energy and the magnon energy [Eq.~\eqref{additional zeeman}] for $(J,D,K)=$ (a) $(1.387,0.09293,2)$, (b) $(0.8876,0.0,4)$, (c) $(1.387,-0.1568,2)$ in meV, which are parameters (2-4) in Fig.~\ref{General two model structure} and ~\ref{General models MOMNC}.
    For three cases, we set $\Delta B$ = 1 T and ten cases of $B$ = 0.1, 1.1, $\cdots$, 8.1, 9.1 T.
    They are plotted along the symmetric points in the reciprocal space connected by a directed red line on the hexagon in (a), as in Fig.~\ref{General two model structure}.
    To indicate the bands, we used solid lines for the lower band and dashed lines for the upper band.
    For each given band, the larger magnetic field is colored with a bright one, and the yellow arrow denotes the direction of increase of the magnetic field for the upper band in each system.}
    \label{figF1}
\end{figure}

We obtain the ratio between the additional magnon energy increase to Zeeman energy and the magnon energy $\Delta E(B,\Delta B)/E(B)$, which is defined by the ratio between the additional magnon energy increase to Zeeman energy and the magnon energy,
\begin{equation}\label{additional zeeman}
\frac{\Delta E(B,\Delta B)}{E(B)}=\frac{E(B+\Delta B)-E(B)-g\mu_B \Delta B}{E(B)},    
\end{equation}
where $E(B)$ is the magnon energy at the magnetic field $B$, and $\Delta B$ is the increase of the external magnetic field.
In Fig.~\ref{figF1}, The ratio for two $J$-$DM$-$K$-$B$ models (a, c) and one $J$-$K$-$B$ model (b), the model parameters are identical to (2-4) in Fig.~\ref{General models MOMNC}, with $\Delta B$ = 1 T and ten different $B$ = 0.1, 1.1, $\cdots$, 8.1, 9.1 T.
We did not plot the $J$-$DM$-$B$ models because $E(B+\Delta B)=E(B)+g\mu_B \Delta B$~\cite{Owerre2016JApplPhys.120.043903,Kim2016PhysRevLett.117.227201}, i.e., $\Delta E(B,\Delta B)/E(B)=0$.
The dashed lines in Fig.~\ref{figF1} indicate the upper band case, of which the difference between $B$ is hardly visible, whereas the solid lines indicate the lower band case whose changes by $B$ increase are directed by the yellow arrow.
From the ratio in Fig.~\ref{figF1}, focusing on the near $\Gamma$ point is enough to include the additional magnon energy increase effect.

\begin{figure}
    \centering
    \includegraphics[width=500pt]{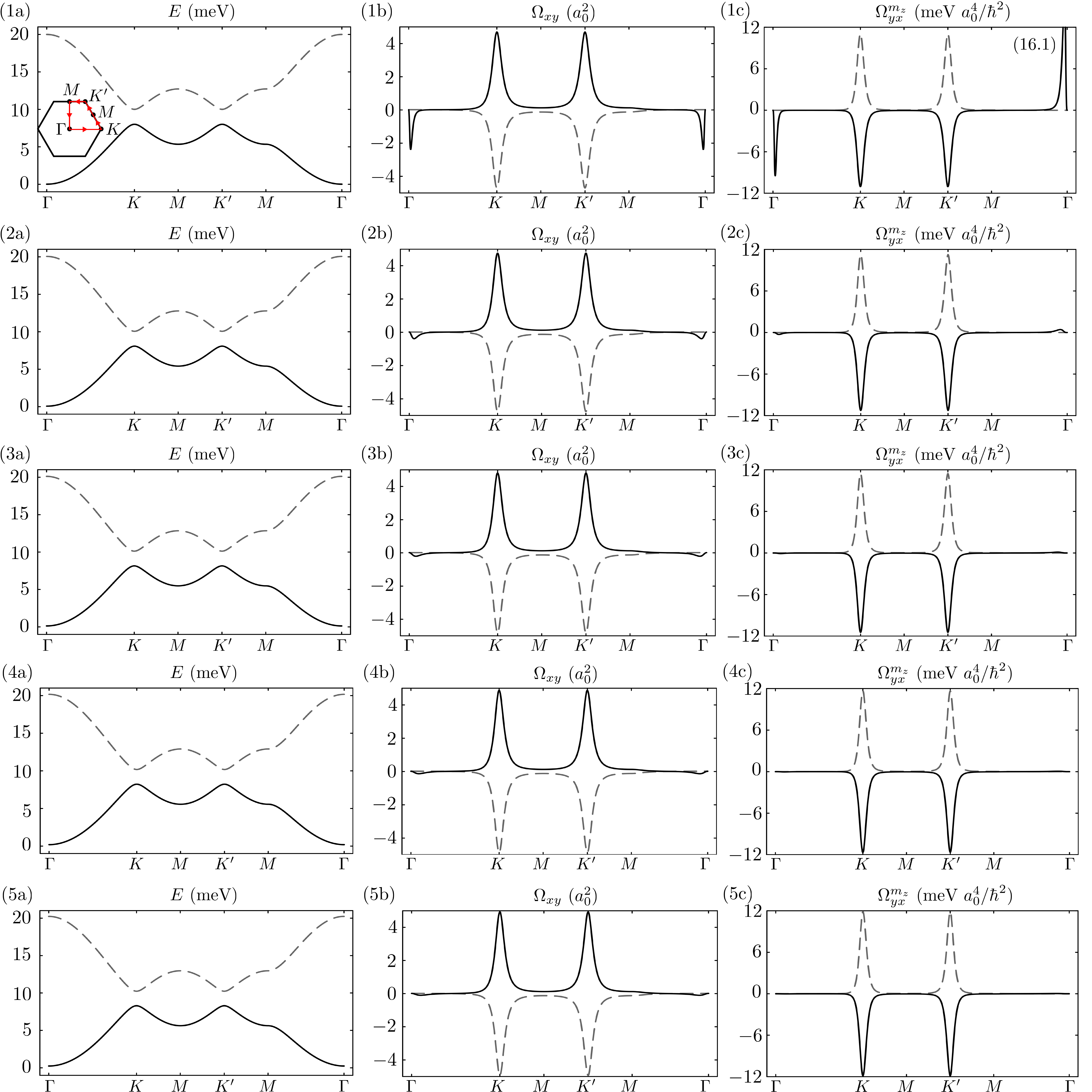} 
    \caption{The numbers from 1 to 5 denote different external magnetic fields $B=$ $0.1$, $0.6$, $1.1$, $1.6$, $2.1$ T respectively.
    For each row, magnon energy band (1-5a), Berry curvature (1-5b), and MOMBC (1-5c) are plotted with parameters $(J,D,K)=(0.8876,0.0,4)$ in meV.
    The number in the parentheses in (1c) denotes the value at the lower band's near $\Gamma$ point peak.
    They are plotted along the symmetric points in the reciprocal space connected by a directed red line on the hexagon in (1a), as in Fig.~\ref{General two model structure}.
    To indicate the bands, we used darker solid lines for the lower band and brighter dashed lines for the upper band.}
    \label{figF2}
\end{figure}

In Fig.~\ref{figF2} (1-5b, 1-5c), we observe that the near $\Gamma$ point peaks of the Berry curvature~\cite{McClarty2018PhysRevB.98.060404} and MOMBC in the $J$-$K$-$B$ model are shortened as the external magnetic field increases, even though the energy band is merely changed in Fig.~\ref{figF2} (1-5a) because the Zeeman interaction energy $g\mu_B B$ of $B=1$ T is 0.116 meV.
The magnitudes of the near $\Gamma$ point peaks, both negative and positive peaks, decrease.

\end{widetext}
\end{appendix}

\end{document}